\documentclass[aps,prx,reprint,preprintnumbers,superscriptaddress,nofootinbib,longbibliography,floatfix]{revtex4-2}
\pdfoutput=1
\usepackage{rotating}
\usepackage{array}
\usepackage{amsmath}
\usepackage[normalem]{ulem}
\usepackage{slashed}
\usepackage{booktabs}
\usepackage[pdftex,table]{xcolor}
\usepackage{units}
\usepackage{xfrac}
\usepackage{mathtools}
\usepackage{empheq}
\usepackage[]{units}
\usepackage{multirow}
\usepackage{amssymb}
\usepackage{url}
\usepackage{comment}
\usepackage{physics}
\usepackage{color,soul}
\usepackage{bbm}
\usepackage[caption=false]{subfig}
\usepackage{hyperref}

\begin{document}

\title{Unbinned Profiled Unfolding}

\author{Jay Chan}
\email{cchan62@wisc.edu}
\affiliation{Department of Physics, University of Wisconsin-Madison, Madison, WI 53706, USA}
\affiliation{Physics Division, Lawrence Berkeley National Laboratory, Berkeley, CA 94720, USA}

\author{Benjamin Nachman}
\email{bpnachman@lbl.gov}
\affiliation{Physics Division, Lawrence Berkeley National Laboratory, Berkeley, CA 94720, USA}
\affiliation{Berkeley Institute for Data Science, University of California, Berkeley, CA 94720, USA}

\date{\today}

\begin{abstract}

Unfolding is an important procedure in particle physics experiments which corrects for detector effects and provides differential cross section measurements that can be used for a number of downstream tasks, such as extracting fundamental physics parameters. Traditionally, unfolding is done by discretizing the target phase space into a finite number of bins and is limited in the number of unfolded variables. Recently, there have been a number of proposals to perform unbinned unfolding with machine learning. However, none of these methods (like most unfolding methods) allow for simultaneously constraining (profiling) nuisance parameters. We propose a new machine learning-based unfolding method that results in an unbinned differential cross section and can profile nuisance parameters.  The machine learning loss function is the full likelihood function, based on binned inputs at detector-level.  We first demonstrate the method with simple Gaussian examples and then show the impact on a simulated Higgs boson cross section measurement.
\end{abstract}

\maketitle

\section{Introduction}

One of the most common analysis goals in particle and nuclear physics is the measurement of differential cross sections.  These quantities encode the rate at which a particular process occurs as a function of certain observables of interest.  From measured cross sections, a number of downstream inference tasks can be performed, including the estimation of fundamental parameters, tuning simulations, and searching for physics beyond the Standard Model.  The key challenge of cross section measurements is correcting the data for detector distortions, a process called deconvolution or \textit{unfolding}.  See Refs.~\cite{Cowan:2002in,Blobel:2203257,doi:10.1002/9783527653416.ch6,Balasubramanian:2019itp} for recent reviews on unfolding and Refs.~\cite{DAgostini:1994fjx,Hocker:1995kb,Schmitt:2012kp} for the most widely-used unfolding algorithms.

Until recently, all cross section measurements were performed with histograms.  In particular, the target spectra and experimental observations were binned and the unfolding problem is recast in the language of linear algebra. That is, one would like to determine the signal strength, defined as the ratio of the observed signal yield to the theoretical prediction, for each bin based on the measurements from experimental observations. This approach comes with the limitation that the binning must be determined beforehand. This makes it difficult to compare measurements with different binning. Furthermore, the optimal binning depends on the downstream inference task.

Modern machine learning (ML) has enabled the creation of unfolding methods that can process unbinned data~\cite{Arratia:2021otl}.  Deep generative models such as Generative Adversarial Networks (GAN)~\cite{Goodfellow:2014:GAN:2969033.2969125,Datta:2018mwd,Bellagente:2019uyp} and Variational Autoencoders (VAE)~\cite{kingma2014autoencoding,Howard:2021pos} produce implicit models that represents the probability density of the unfolded result and allow to sample from the probability density. Methods based on Normalizing Flows (NF)~\cite{10.5555/3045118.3045281,Bellagente:2020piv,Vandegar:2020yvw,Backes:2022vmn} allow for both sampling and density estimation.  In contrast, the classifier-based method OmniFold Refs.~\cite{Andreassen:2019cjw,Andreassen:2021zzk} iteratively reweights a simulated dataset.   A summary of machine learning-based unfolding methods can be found in Ref.~\cite{Arratia:2021otl} and recent applications of these techniques (in particular, of OmniFold) to experimental data are presented in Refs.~\cite{H1:2021wkz,H1prelim-22-031,H1prelim-22-034,LHCb:2022rky}.  While powerful, none of these approaches can simultaneously estimate cross sections and fit (nuisance) parameters.  This can be a significant shortcoming when the phase space region being probed has non-trivial constraining power for systematic uncertainties.

Unfolding methods that can also profile have been proposed.  One possibility is to treat the cross section in each region of particle-level phase space (i.e. in a histogram bin) as a free parameter and then perform a likelihood fit as for any set of parameters of interest and nuisance parameters.  For example, this is the setup of the the Simplified Template Cross Section (STXS) (e.g. Refs.~\cite{LHCHiggsCrossSectionWorkingGroup:2016ypw, Andersen:2016qtm, Berger:2019wnu, Amoroso:2020lgh}) measurements for Higgs boson kinematic properties. Another possibility is Fully Bayesian Unfolding (FBU) \cite{https://doi.org/10.48550/arxiv.1201.4612}, which samples from the posterior probability over the cross section in each bin of the particle-level phase space and over the nuisance parameters.  All of these methods require binning.

In this paper, we propose a new machine learning-based unfolding method that is both unbinned at particle level and can profile, referred to as Unbinned Profiled Unfolding (UPU). UPU reuses all the standard techniques used in binned maximum likelihood unfolding and combines them with ML methods that allow for unbinned unfolding. Specifically, we use the binned maximum likelihood at detector level as the metric to optimize the unfolding, while the unfolding takes unbinned particle-level simulations as inputs.

The rest of this paper is organized as follows. In Sec.~\ref{sec:UPU}, we describe the procedure and implementation details of UPU. We then present simple Gaussian examples to demonstrate the usage of UPU in Sec.~\ref{sec:gaussian}. In Sec.~\ref{sec:higgs}, we apply UPU to a simulated Higgs boson cross section measurement at the Large Hadron Collider (LHC). The conclusions and outlook are then given in Sec.~\ref{sec:conclusion}.

\section{Unbinned Profiled Unfolding}
\label{sec:UPU}

\subsection{Statistical Setup}

UPU generalizes binned maximum likelihood unfolding to the unbinned case.  Binned maximum likelihood unfolding can be described by the following optimization setup:

\begin{align}
\label{eq:binned}
    (\hat{k},\hat{\theta})=\text{argmax}_{(k,\theta)}\Pr(m|k,\theta)\,p_0(\theta)\,,
\end{align}
where $m\in\mathbb{R}^{N_m}$ is a vector representing the counts in each of the $N_m$ bins at detector level, $k\in\mathbb{R}^{N_k}$ is a vector representing the counts in each of the $N_k$ bins at particle level (usually $N_m\geq N_k$), $\theta$ are the nuisance parameters, and $p_0$ is the prior on $\theta$.  Our idea is to keep the structure of Eq.~\ref{eq:binned}, but replace $k$ with an unbinned estimator of the particle-level spectrum.  Suppose that the particle-level phase space is\footnote{Assuming the space is suitably standardized to remove units.} $\mathbb{R}^N$ and let\footnote{We will use $[\cdot]$ to denote the parameters of the function and $(\cdot)$ to denote the inputs of the function, e.g. $f[\theta](x)$ is a functional in $\theta$ and a function in $x$.} $\tau[\omega]\in \mathbb{R}^{\mathbb{R}^N}$ parameterize the probability density over this space for parameters $\omega$. The goal of UPU is then to optimize

\begin{align}
\label{eq:unbinned1}
    (\hat{\omega},\hat{\theta})= \text{argmax}_{(\omega,\theta)}\Pr(m|\tau[\omega],\theta)\,p_0(\theta)\,,
\end{align}
where the final result would be given by $\tau[\hat{\omega}]$.  The challenge with the construction in Eq.~\ref{eq:unbinned1} is that for a given truth spectrum $\tau[\omega]$, we need to know the predicted detector-level distribution.  In the binned case, this is readily computed by multiplying $k$ by the response matrix $R_{ij}=\Pr(\text{measure in bin $i$}|\text{truth is bin $j$})$.  When the truth are unbinned, we need the full detector response. This is never known analytically and would be challenging to estimate numerically with a surrogate density estimator\footnote{Note that Eq.~\ref{eq:unbinned1} is a probability distribution over probability distributions so building it from the per-event detector response is non-trivial.}.  To address this challenge, we make use of the fact that simulated events come in pairs, with a matching between particle-level and detector-level events.  Instead of estimating $\tau$ directly, we use a fixed simulation (with particle-level spectrum $\tau[\omega_0]$) and then learn a reweighting function $w_0[\lambda]$ to estimate the likelihood ratio between the unfolded result and the fixed simulation at particle level.  Schematically:
\begin{align}
\label{eq:unbinned}
    (\hat{\lambda},\hat{\theta})= \text{argmax}_{(\lambda,\theta)}\Pr(m|\tau [\omega_0]w_0[\lambda],\theta)\,p_0(\theta)\,,
\end{align}
where in practice, we only have samples from $\tau[\omega_0]$ and $w_0$ is a surrogate model.  The number of predicted events in a given bin $i$ is then a sum over weights $w_0[\hat{\lambda}]$ (evaluated at particle-level) for simulated events with a detector-level value in bin $i$.  The probability over values $m$ is then a product over Poisson probability mass functions, since the bins are statistically independent.  The fact that the probability mass is known is crucial and means that UPU does not readily generalize the case where the detector-level phase space is also unbinned. This is the case for OmniFold, which also uses reweighting at particle-level. In contrast to UPU, OmniFold uses an Expectation-Maximization-type algorithm to iteratively converge to the maximum likelihood estimate and does not currently allow for mixed implicit/explicit likelihood constraints (as is needed for nuisance parameters).

\begin{table*}[htbp]
  \begin{center}
    \caption{Summary of Gaussian example data sets.}
    \label{tab:dataset}
    \small
    \begin{tabular}{cccc}
      \toprule
     Data set & Parameters & Number of events & Purpose \\
      \midrule
      $D_\mathrm{sim}^{1.0}$ & $\mu=0$, $\sigma=1$ and $\epsilon=1$ & 200,000 & Nominal simulation \\
      $D_\mathrm{obs}$ & $\mu=0.8$, $\sigma=1$ and $\epsilon=1.2$ & 100,000 & Observed data \\
      $D_\mathrm{sim}^*$ & $\mu=0$, $\sigma=1$ and $\epsilon=\left(0.2, 1.8\right)$ & 200,000 & Train $w_1$ \\
      $D_\mathrm{sim}^{1.2}$ & $\mu=0$, $\sigma=1$ and $\epsilon=1.2$ & 100,000 & Validate $w_1$ \\
      \bottomrule
    \end{tabular}
  \end{center}
\end{table*}

\subsection{Machine Learning Approach}

For particle-level features $T$ and detector-level features $R$, the main goal is to train the likelihood ratio estimator $w_0\left(T\right)$, which reweights the simulated particle-level spectrum.  In the absence of profiling, this corresponds to the following loss function:

\begin{align}
\label{eq:lossnoprofile}
    L &= \prod_{i=1}^{n_\text{bins}}\Pr(n_i\Bigg|\sum_{j=1}^{n_\text{MC}}w_0(T_j)\mathbb{I}_i(R_j))\,,
\end{align}
where $n_i$ is the number of observed events in bin $i$, $n_\text{MC}$ is the number of simulated events, and $\mathbb{I}_i(\cdot)$ is the indicator function that is one when $\cdot$ is in bin $i$ and zero otherwise.  When $w_0$ is parameterized as a neural network (see Sec.~\ref{sec:implement}), then the logarithm of Eq.~\ref{eq:lossnoprofile} is used for training:

\begin{align}
\label{eq:lossnoprofile2}
    &\log L = \\\nonumber
    &\hspace{2mm}\sum_{i=1}^{n_\text{bins}}\left[n_i\log\left(\sum_{j=1}^{n_\text{MC}}w_0(T_j)\mathbb{I}_i(R_j)\right)-\sum_{i=1}^{n_\text{MC}}w_0(T_j)\mathbb{I}_i(R_j)\right]\,,
\end{align}
where we have dropped constants that do not affect the optimization.  Experimental nuisance parameters modify the predicted counts in a particular bin given the particle-level counts.  We account for these effects with a second reweighting function:

\begin{align}
    w_1(R|T,\theta)=\frac{p_{\theta}(R|T)}{p_{\theta_0}(R|T)}\,,
\end{align}
where $p_\theta(R|T)$ is the conditional probability density of $R$ given $T$ with nuisance parameters $\theta$.  Importantly, $w_1$ does not modify the target particle level distribution.  Incorporating $w_1$ into the log likelihood results in the full loss function:


\begin{equation}
\begin{split}
    \log L&=\sum_{i=1}^{n_\text{bins}}\Biggl[n_i\log\left(\sum_{j=1}^{n_\text{MC}}w_0(T_j)w_1(R_j|T_j,\theta)\mathbb{I}_i(R_j)\right) \\
     &-\sum_{j=1}^{n_\text{MC}}w_0(T_j)w_1(R_j|T_j,\theta)\mathbb{I}_i(R_j)\Biggr]+\log p_0(\theta)\,.
\end{split}
\label{eq:lllall}
\end{equation}

Since $w_1$ does not depend on the particle-level spectrum, it can be estimated prior to the final fit and only the parameters of $w_0$ and the value(s) of $\theta$ are allowed to float when optimizing Eq.~\ref{eq:lllall}.

\subsection{Machine Learning Implementation}
\label{sec:implement}

In our subsequent case studies, the reweighting functions $w_0$ and $w_1$ are parametrized with neural networks.  The $w_0$ function is only constrained to be non-negative and so we choose it to be the exponential of a neural network.

The pre-training of $w_1$ requires neural conditional reweighting~\cite{2107.08979}, as a likelihood ratio in $R$ conditioned on $T$ and parameterized in $\theta$.  While there are multiple ways of approximating conditional likelihood ratios, the one we found to be the most stable for the examples we have studied for UPU is the product approach:

\begin{align}
    w_1(R|T,\theta)=\left(\frac{p_\theta(R,T)}{p_{\theta_0}(R,T)}\right)\left(\frac{p_{\theta_0}(T)}{p_{\theta}(T)}\right)\,,
\end{align}
where the two terms on the righthand side are separately estimated and then their product is $w_1$.  For a single feature $T$, a likelihood ratio between samples drawn from a probability density $p$ and samples drawn from a probability density $q$ is estimated using the fact that machine learning-classifiers approximate monotonic transformations of likelihood ratios (see e.g. Ref.~\cite{hastie01statisticallearning,sugiyama_suzuki_kanamori_2012}).  In particular, we use the standard binary cross entropy loss function 

\begin{align}
L_\text{BCE}[f]=-\sum_{Y\sim p}\log(f(Y))-\sum_{Y\sim q}\log(1-f(Y))\,,
\end{align}
and then the likelihood ratio is estimated as $f/(1-f)$.  The last layer of the $f$ networks are sigmoids in order to constrain their range to be between 0 and 1.  The function $f$ is additionally trained to be parameterized in $\theta$ by training with pairs $(Y,\Theta)$ instead of just $Y$, where $\Theta$ is a random variable corresponding to values $\theta$ sampled from a prior.  We will use a uniform prior when training the parameterized classifiers.  

All neural networks are implemented using PyTorch \cite{NEURIPS2019_9015} and optimized with Adam \cite{https://doi.org/10.48550/arxiv.1412.6980} with a learning rate of 0.001 and consist of three hidden layers with 50 nodes per layer.  All intermediate layers use ReLU activation functions.  Each network is trained for 10,000 epochs with early stopping using a patience of 10.  The $w_1$ training uses a batch size of 100,000.  The $w_0$ network is simultaneously optimized with $\theta$ and uses a batch size that is the full dataset, which corresponds to performing the fit in Eq.~\ref{eq:lllall} over all the data.
All neural networks are implemented using PyTorch \cite{NEURIPS2019_9015} and optimized with Adam \cite{https://doi.org/10.48550/arxiv.1412.6980} with a learning rate of 0.001 and consist of three hidden layers with 50 nodes per layer.  All intermediate layers use ReLU activation functions.  Each network is trained for 10,000 epochs with early stopping using a patience of 10.  The $w_1$ training uses a batch size of 100,000.  The $w_0$ network is simultaneously optimized with $\theta$ and uses a batch size that is the full dataset, which corresponds to performing the fit in Eq.~\ref{eq:lllall} over all the data.

\section{Gaussian Example}
\label{sec:gaussian}

We now demonstrate the proposed method with a simple numerical example. Here, each data set represents a one-dimension Gaussian distribution in the particle level and a two-dimension distribution in the detector level. The particle-level Gaussian random variable $T$ is described by mean $\mu$ and standard deviation $\sigma$, while the detector-level variables are given by
\begin{align}
    R & = T + Z,\\
    R^* & = T + Z^*,
\end{align}
where $Z$ ($Z^*$) is a Gaussian random variable with mean $\beta$ ($\beta^*$) and standard deviation $\epsilon$ ($\epsilon^*$). $\epsilon$ is considered to be the only nuisance parameter, and $\beta$, $\beta^*$ are fixed to 0, and $\epsilon^*$ is fixed to 1. In this case, the nuisance parameter $\epsilon$ only has effect on the $R$ spectrum and the $R^*$ spectrum depends purely on the particle-level spectrum $T$. This setup is thus sensitive to both the effect of $w_0$ and that of $w_1$ \footnote{An ill-defined example is shown in App.~\ref{sec:ssec:gaussian1D}, where the considered detector-level observable, a one-dimension Gaussian distribution, is not able to distinguish between effects from particle level and effects from detector level with $\theta$. This limitation also exists in the standard binned maximum likelihood unfolding, as shown in App.~\ref{app:bmlu}}.

Three data sets are prepared for the full training procedure. As summarized in Tab.~\ref{tab:dataset}, the first data set $D_\mathrm{sim}^{1.0}$ is used as the nominal simulation sample, which contains 200,000 events with $\mu=0$, $\sigma=1$ and $\epsilon=1$. The second data set $D_\mathrm{obs}$ is used as the observed data, which contains 100,000 events with $\mu=0.8$, $\sigma=1$ and $\epsilon=1.2$. To train the $w_1$ reweighter, the third data set $D_\mathrm{sim}^*$, which contains 200,000 events with $\mu=0$, $\sigma=1$ and $\epsilon$ uniformly distributed from 0.2 to 1.8, is prepared and used as the simulation with systematic variations. In addition, another data set $D_\mathrm{sim}^{1.2}$ of 100,000 events with $\mu=0$, $\sigma=1$ and $\epsilon=1.2$ is produced for validating the $w_1$ reweighter. All data sets used in the training procedure are split to 50\% for training and 50\% for validating.

\begin{figure*}[htbp]
\begin{center}
\includegraphics[width=0.43\textwidth]{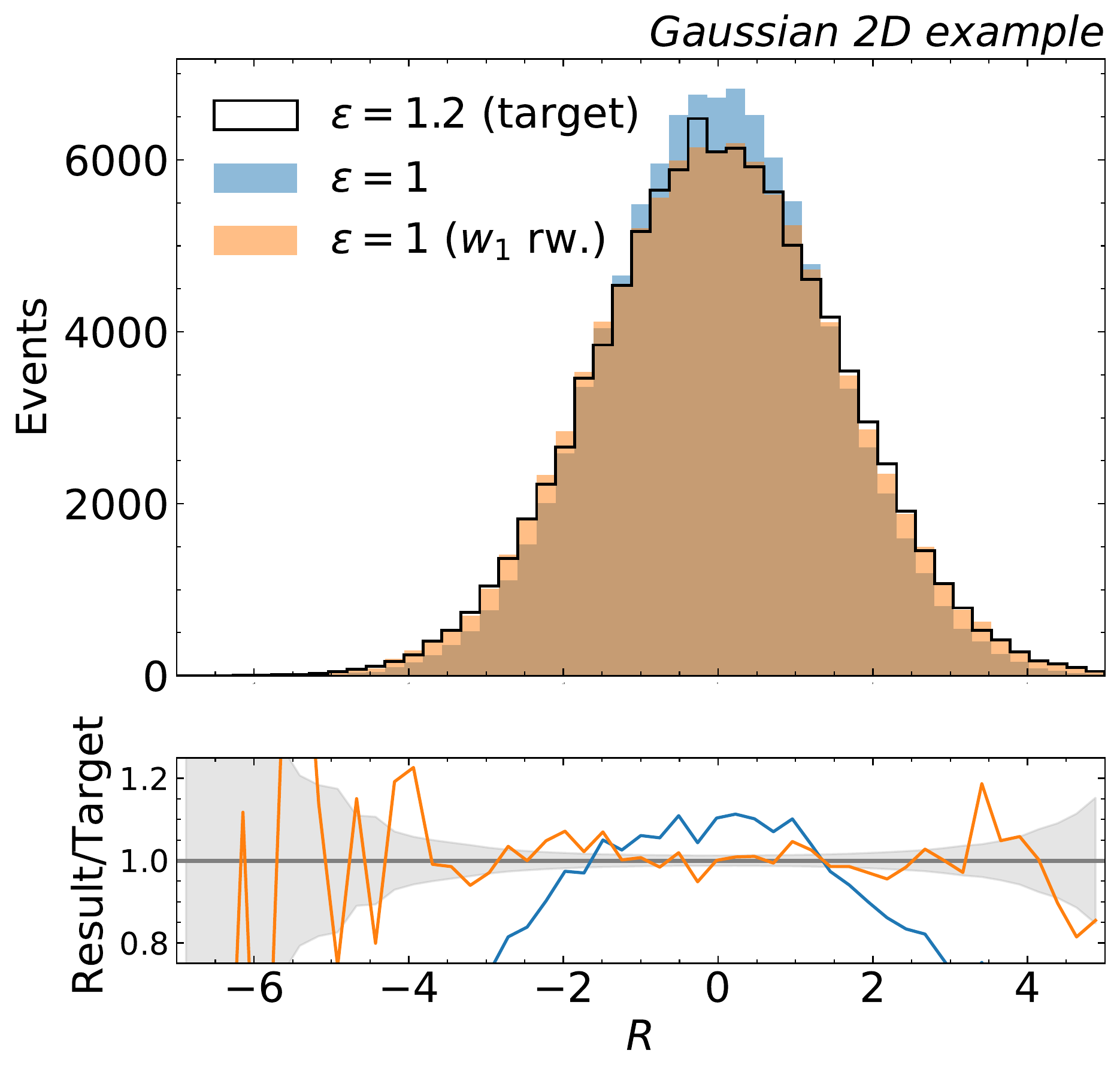}
\includegraphics[width=0.43\textwidth]{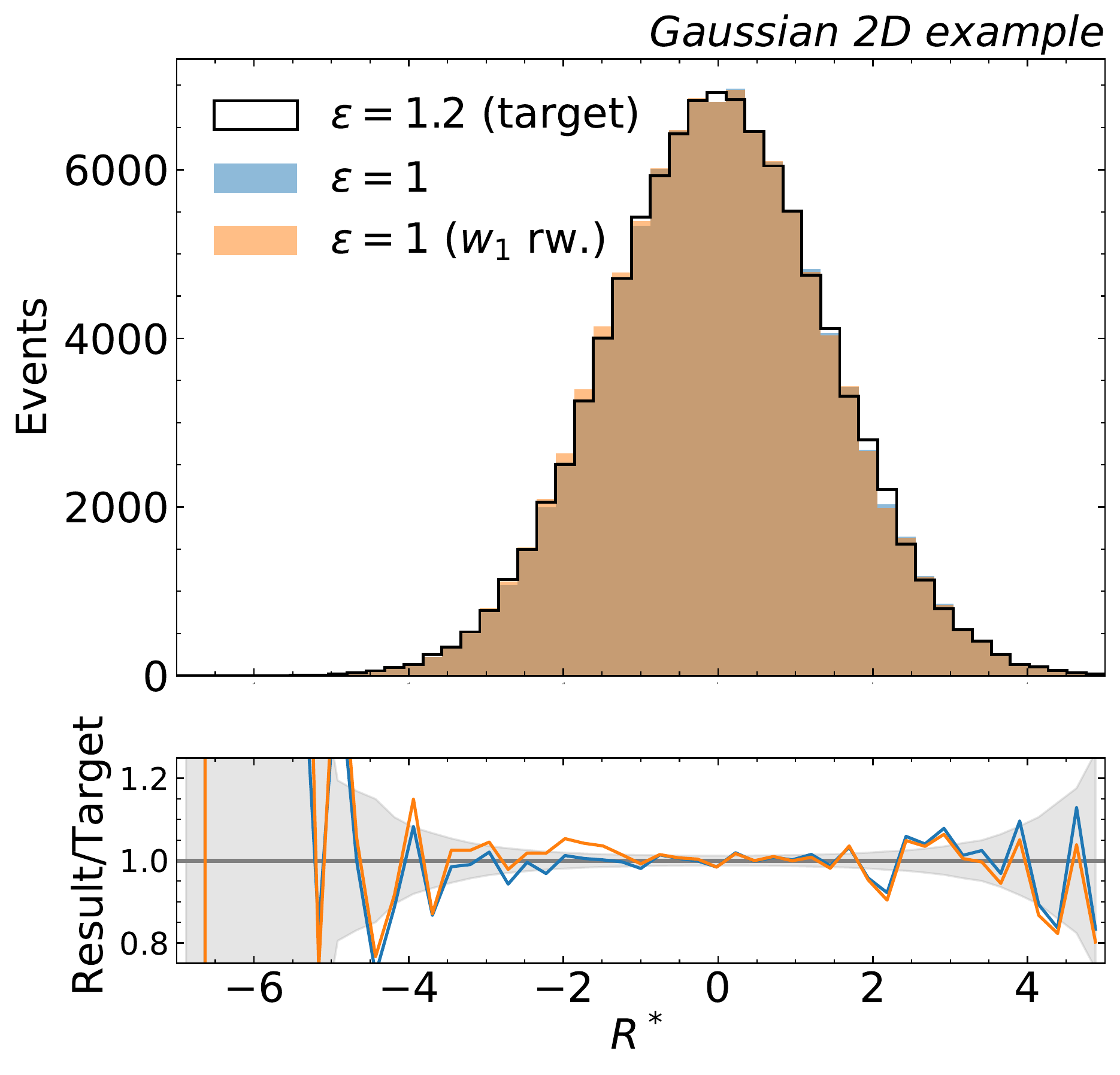}
\caption{Gaussian 2D example: the nominal detector-level spectra $R$ (left) and $R^*$ (right) with $\epsilon = 1$ reweighted by the trained $w_1$ conditioned at $\epsilon = 1.2$ and compared to the spectra with $\epsilon = 1.2$. The shaded band in the bottom panel represents the data statistical uncertainty, which is estimated as $1/\sqrt{n}$, where $n$ is the number of $D_\mathrm{sim}^{1.2}$ events in a given bin.}
\label{fig:gaussian_2D_w1}
\end{center}
\end{figure*}

A $w_1$ reweighter is trained to reweight $D_\mathrm{sim}^{1.0}$ to $D_\mathrm{sim}^*$. The trained $w_1$ is tested with the nominal $R$ and $R^*$ spectra ($D_\mathrm{sim}^{1.0}$) reweighted to $\epsilon = 1.2$ and compared to the $R$ and $R^*$ spectra with $\epsilon = 1.2$. As shown in Fig. \ref{fig:gaussian_2D_w1}, the trained $w_1$ reweighter has learned to reweight the nominal $R$ spectrum to match the $R$ spectrum with $\epsilon$ at 1.2, and $R^*$ is independent of the $w_1$ reweighter.

\begin{figure*}[htbp]
\begin{center}
\includegraphics[width=0.43\textwidth]{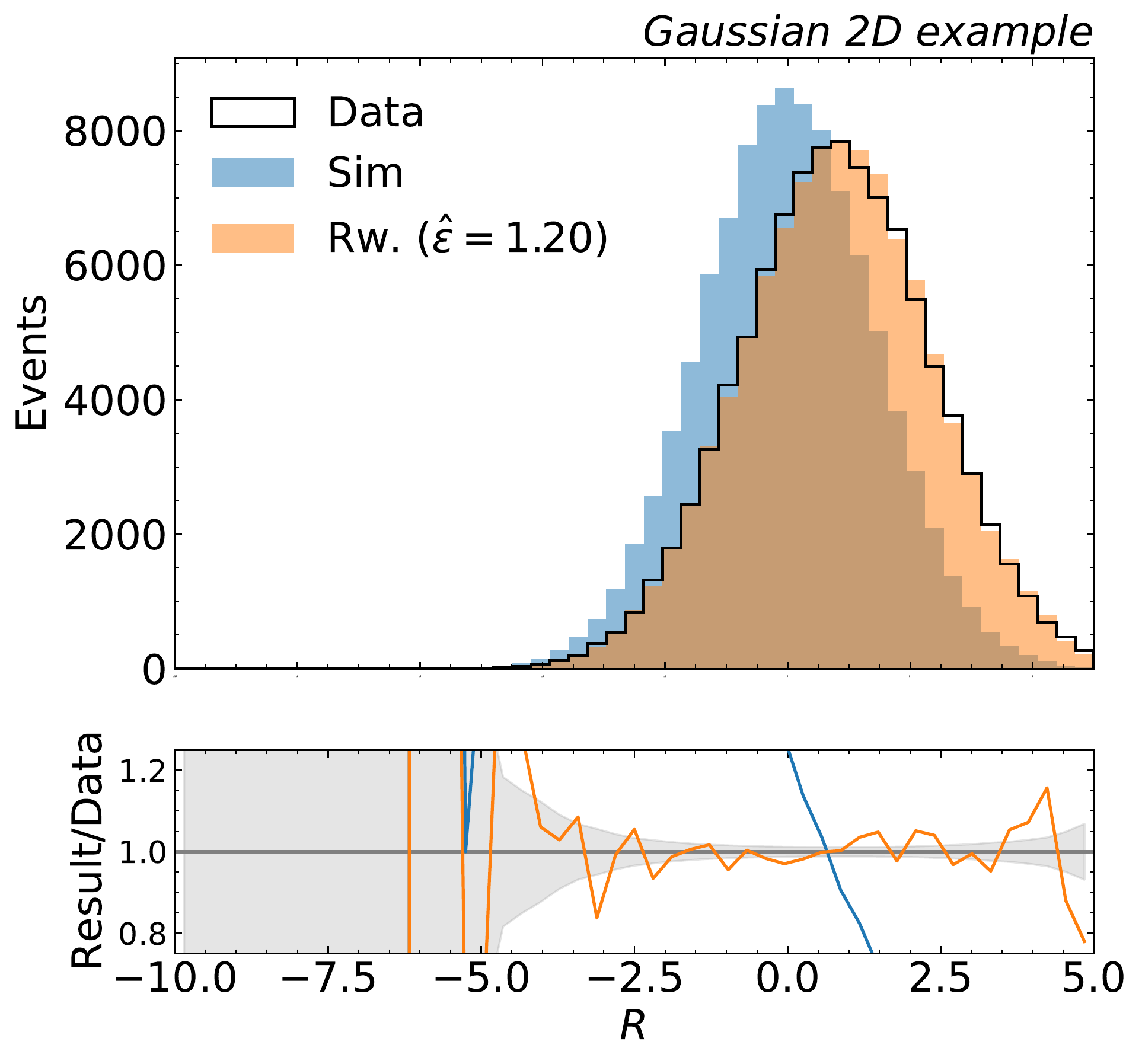}
\includegraphics[width=0.43\textwidth]{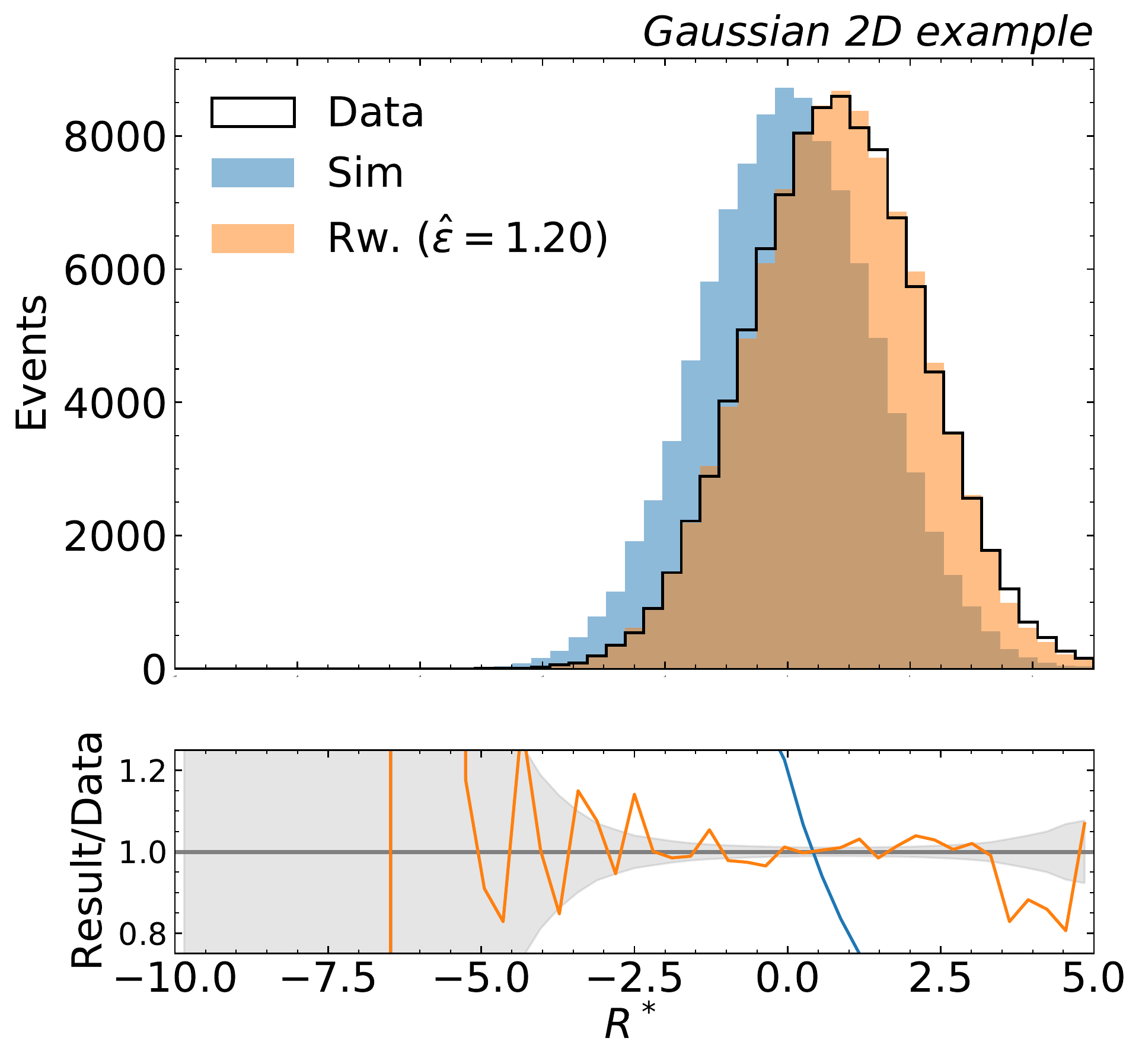}\\
\includegraphics[width=0.43\textwidth]{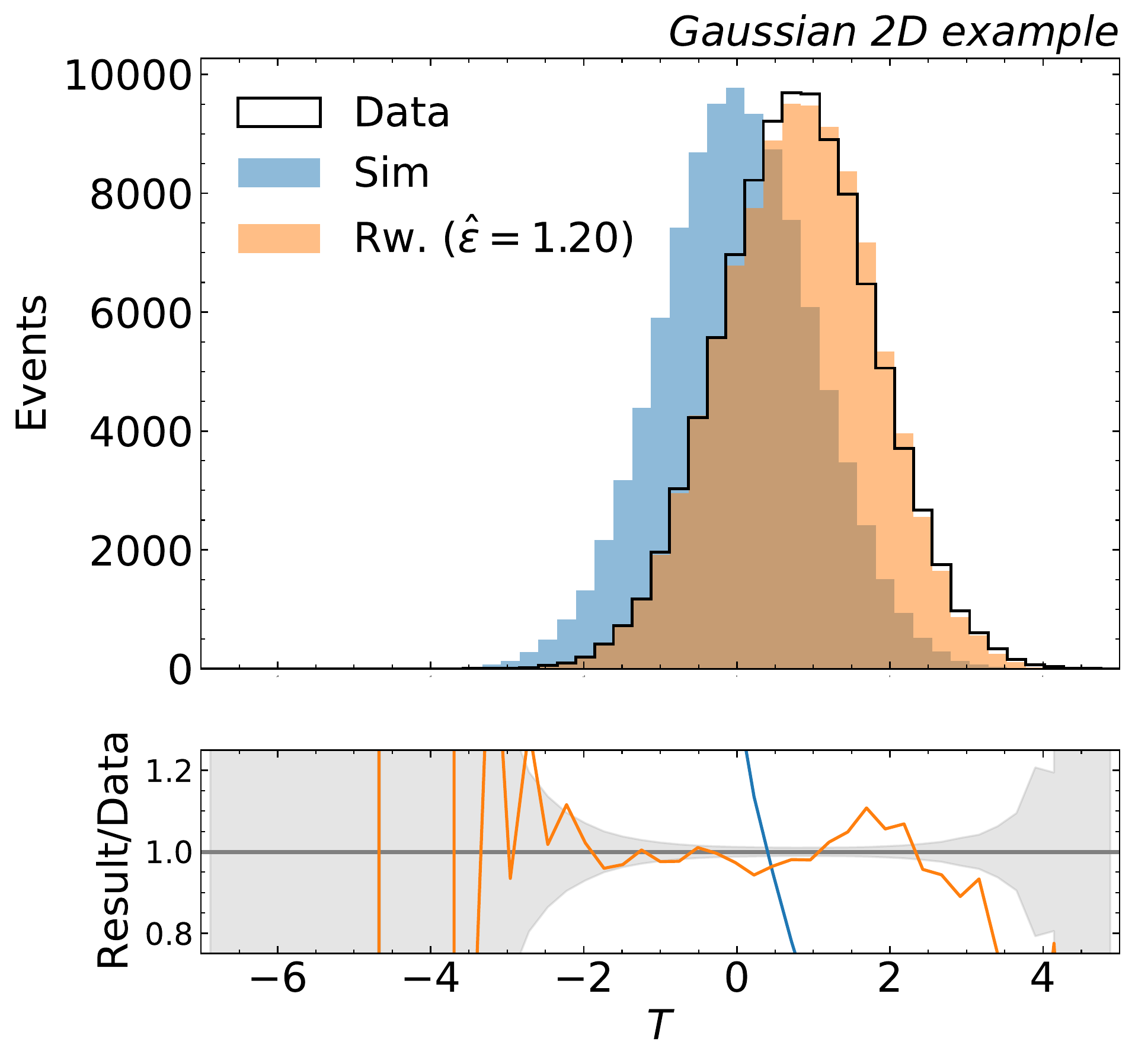}
\caption{Gaussian 2D example: results of the $w_0$ optimization. The nuisance parameter $\epsilon$ is optimized simultaneously with $w_0$ with the prior constraint set to 80\%. The fitted $\epsilon$ is $1.20 \pm 0.004$. (Top-left) The detector-level spectrum $R$ of the simulation template $D_\mathrm{sim}$ reweighted by the trained $w_0 \times w_1$, compared to the $R$ spectrum of the observed data $D_\mathrm{obs}$. (Top-right) The detector-level spectrum $R^\prime$ of the simulation template $D_\mathrm{sim}$ reweighted by the trained $w_0 \times w_1$, compared to the $R^*$ spectrum of the observed data $D_\mathrm{obs}$. (Bottom) The particle-level spectrum $T$ of the simulation template $D_\mathrm{sim}$ reweighted by the trained $w_0$, compared to the $T$ spectrum of the observed data $D_\mathrm{obs}$. The shaded band in the bottom panel represents the data statistical uncertainty, which is estimated as $1/\sqrt{n}$, where $n$ is the number of observed events in a given bin.}
\label{fig:gaussian2D_w0_epsilon_0.8}
\end{center}
\end{figure*}

Based on the trained $w_1$ reweighter, a $w_0$ reweighter and the nuisance parameter $\epsilon$ are optimized simultaneously using $D_\mathrm{sim}$ as the simulation template with $D_\mathrm{obs}$ as the observed data used in Eq.~\ref{eq:lllall}. The prior in the penalty term in Eq.~\ref{eq:lllall} is configured with an uncertainty of 80\%. The fitted $\epsilon$ is $1.20 \pm 0.004$ \footnote{The fitted value is averaged over five different $w_0$ reweighters which are trained in the same way, but with different random initializations. The standard deviation of the fitted values is taken as the error.} (correct value is 1.2). As shown in Fig. \ref{fig:gaussian2D_w0_epsilon_0.8}, the reweighted spectra match well with observed data in both detector and particle level. For more realistic uncertainties (so long as the simulation is close to the right answer), the fidelity is even better.


\section{Higgs Boson Cross Section}
\label{sec:higgs}

We now demonstrate the unfolding method in a physics case --- a Higgs boson cross section measurement. Here, we focus on the di-photon decay channel of the Higgs boson. The goal is then to measure the transverse momentum spectrum of the Higgs boson $p^\mathrm{T}_H$ using the transverse momentum of the di-photon system $p^\mathrm{T}_{\gamma\gamma}$ at detector level. The photon resolution $\epsilon_{\gamma}$ is considered as a nuisance parameter. In this case, the $p^\mathrm{T}_{\gamma\gamma}$ spectrum is minimally affected by $\epsilon_{\gamma}$. Therefore, we also consider the invariant mass spectrum of the di-photon system $m_{\gamma\gamma}$ at detector level, which is highly sensitive to $\epsilon_{\gamma}$. In addition, In order to have a large spectrum difference between different data sets for demonstration purpose, we consider only events that contain at least two reconstructed jets, where the leading-order (LO) calculation would significantly differ from next-to-leading-order calculation (NLO)

Similar to the Gaussian examples, we prepare the following data sets:

\begin{itemize}
    \item $D_\mathrm{obs}$: used as the observed data.
    \item $D_\mathrm{sim}^{1.0}$: used as the nominal simulation sample.
    \item $D_\mathrm{sim}^{1.2}$: used as the simulation sample with a systematic variation.
    \item $D_\mathrm{sim}^{*}$: simulation sample with various $\epsilon_\gamma$ values for training the $w_1$ reweighter.
\end{itemize}

$D_\mathrm{obs}$ is generated at NLO using the \textsc{Powheg}\textsc{Box} program~\cite{Oleari:2010nx, Alioli:2008tz}, while the rest are generated at LO using \textsc{Mad}\textsc{Graph}5\_aMC@LO v2.6.5~\cite{Alwall:2014hca}. For all samples, the parton-level events are processed by \textsc{Pythia} 8.235~\cite{Sjostrand:2006za,Sjostrand:2014zea} for the Higgs decay, the parton shower, hadronization, and the underlying event. The detector simulation is based on \textsc{Delphes} 3.5.0~\cite{deFavereau:2013fsa} with detector response modified from the default ATLAS detector card. For both $D_\mathrm{obs}$ and $D_\mathrm{sim}^{1.2}$, the photon resolution $\epsilon$ is multiplied by a factor of 1.2. For $D_\mathrm{sim}^{*}$, the multiplier of $\epsilon$ is uniformly scanned between 0.5 and 1.5 with a step size of 0.01. $D_\mathrm{sim}^{1.0}$ uses the default ATLAS detector card. 

\begin{figure*}[htbp]
\begin{center}
\includegraphics[width=0.43\textwidth]{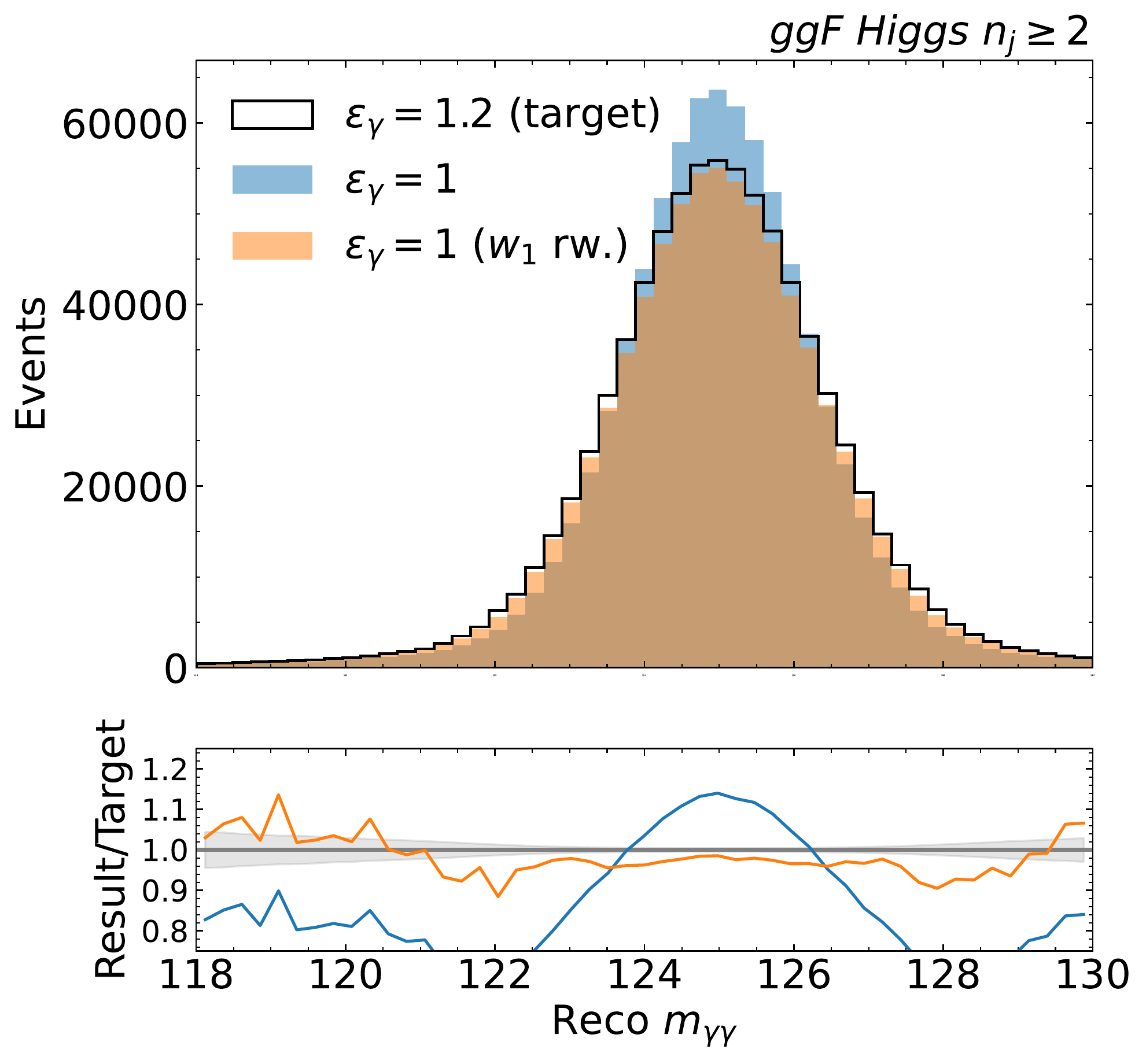}
\includegraphics[width=0.43\textwidth]{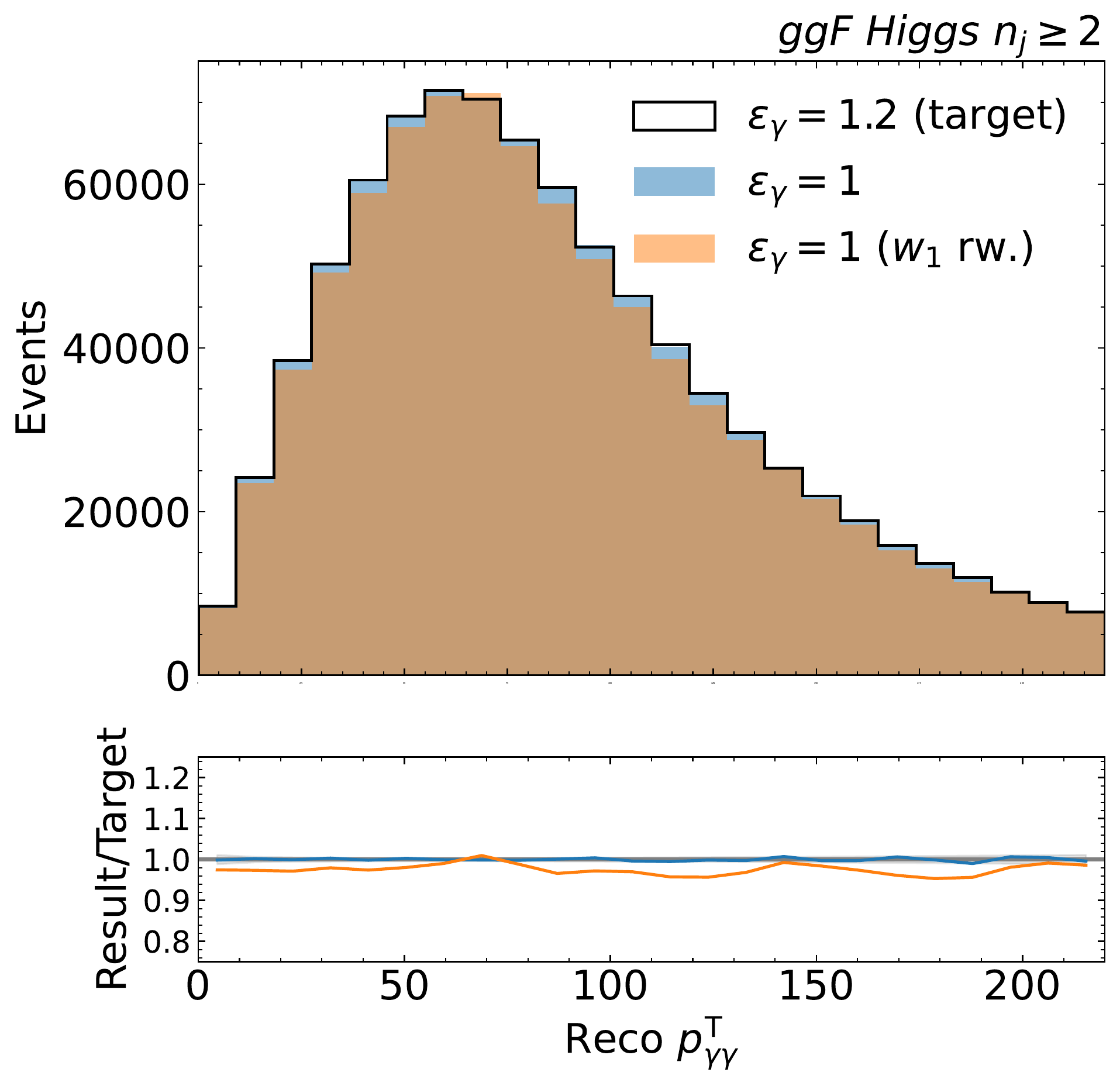}
\caption{Higgs boson cross section: the nominal detector-level spectra $m_{\gamma\gamma}$ (left) and $p^\mathrm{T}_{\gamma\gamma}$ (right) with $\epsilon_\gamma = 1$ reweighted by the trained $w_1$ conditioned at $\epsilon_\gamma = 1.2$ and compared to the spectra with $\epsilon_\gamma = 1.2$. The shaded band in the bottom panel represents the data statistical uncertainty, which is estimated as $1/\sqrt{n}$, where $n$ is the number of $D_\mathrm{sim}^{1.2}$ events in a given bin.}
\label{fig:higgs_w1}
\end{center}
\end{figure*}

Each of the spectra of particle-level $p^\mathrm{T}_{\gamma\gamma}$, detector-level $p^\mathrm{T}_{\gamma\gamma}$ and detector-level $m_{\gamma\gamma}$ is standardized to the spectrum with a mean of 0 and a standard deviation of 1 before being passed to the neural networks. A $w_1$ reweighter is then trained to reweight $D_\mathrm{sim}^{1.0}$ to $D_\mathrm{sim}^*$. The trained $w_1$ is tested with the nominal detector level $p^\mathrm{T}_{\gamma\gamma}$ and $m_{\gamma\gamma}$ spectra ($D_\mathrm{sim}^{1.0}$) reweighted to $\epsilon_{\gamma}=1.2$ and compared to the detector level $p^\mathrm{T}_{\gamma\gamma}$ and $m_{\gamma\gamma}$ spectra with $\epsilon_{\gamma} = 1.2$. As shown in Fig. \ref{fig:higgs_w1}, the trained $w_1$ reweighter has learned to reweight the nominal detector level $m_{\gamma\gamma}$ spectrum to match the detector level $m_{\gamma\gamma}$ spectrum with $\epsilon_{\gamma}$ at 1.2, and the detector level $p^\mathrm{T}_{\gamma\gamma}$ variable is independent of the $w_1$ reweighter.

\begin{figure*}[htbp]
\begin{center}
\includegraphics[width=0.43\textwidth]{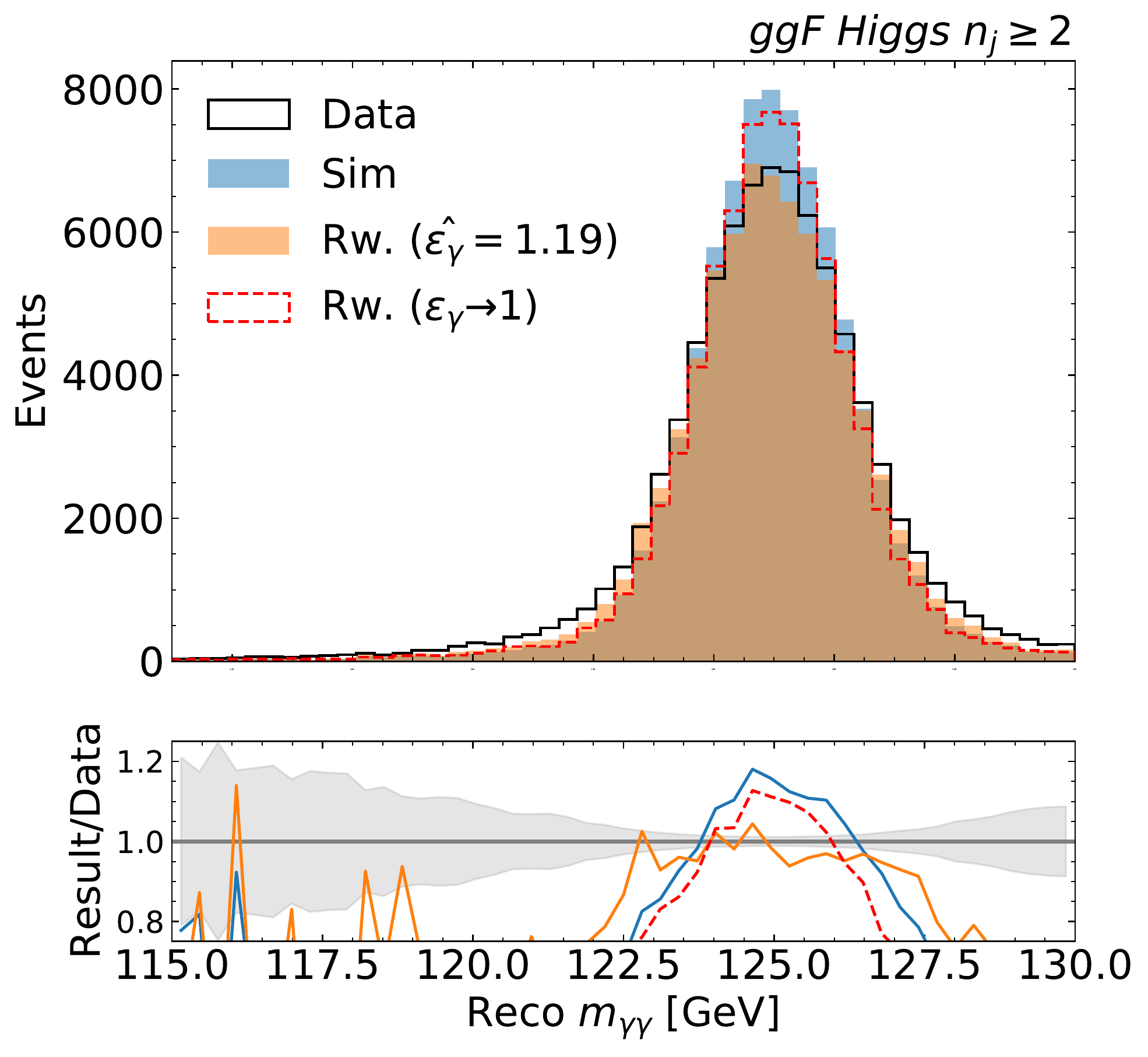}
\includegraphics[width=0.43\textwidth]{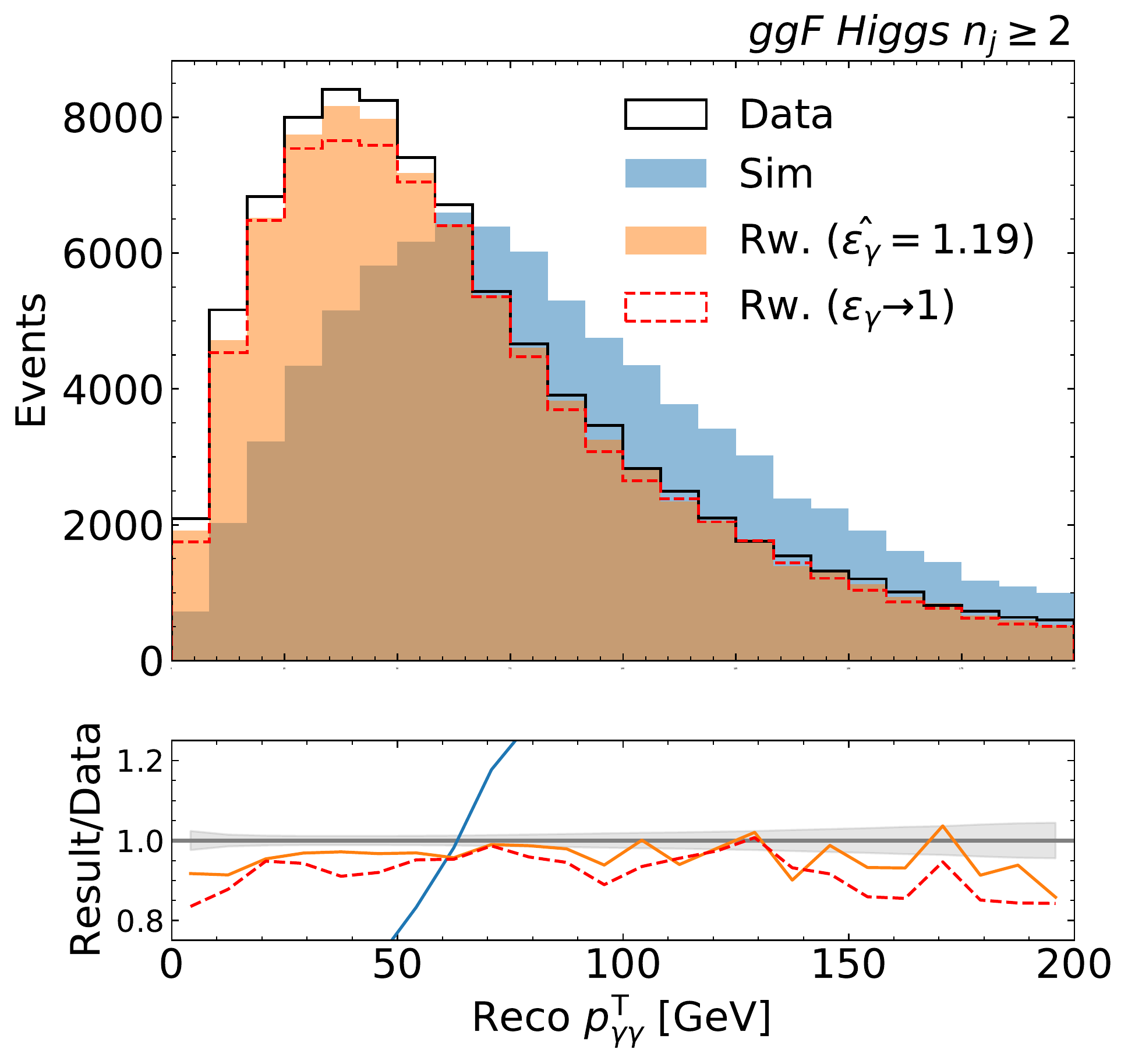}\\
\includegraphics[width=0.43\textwidth]{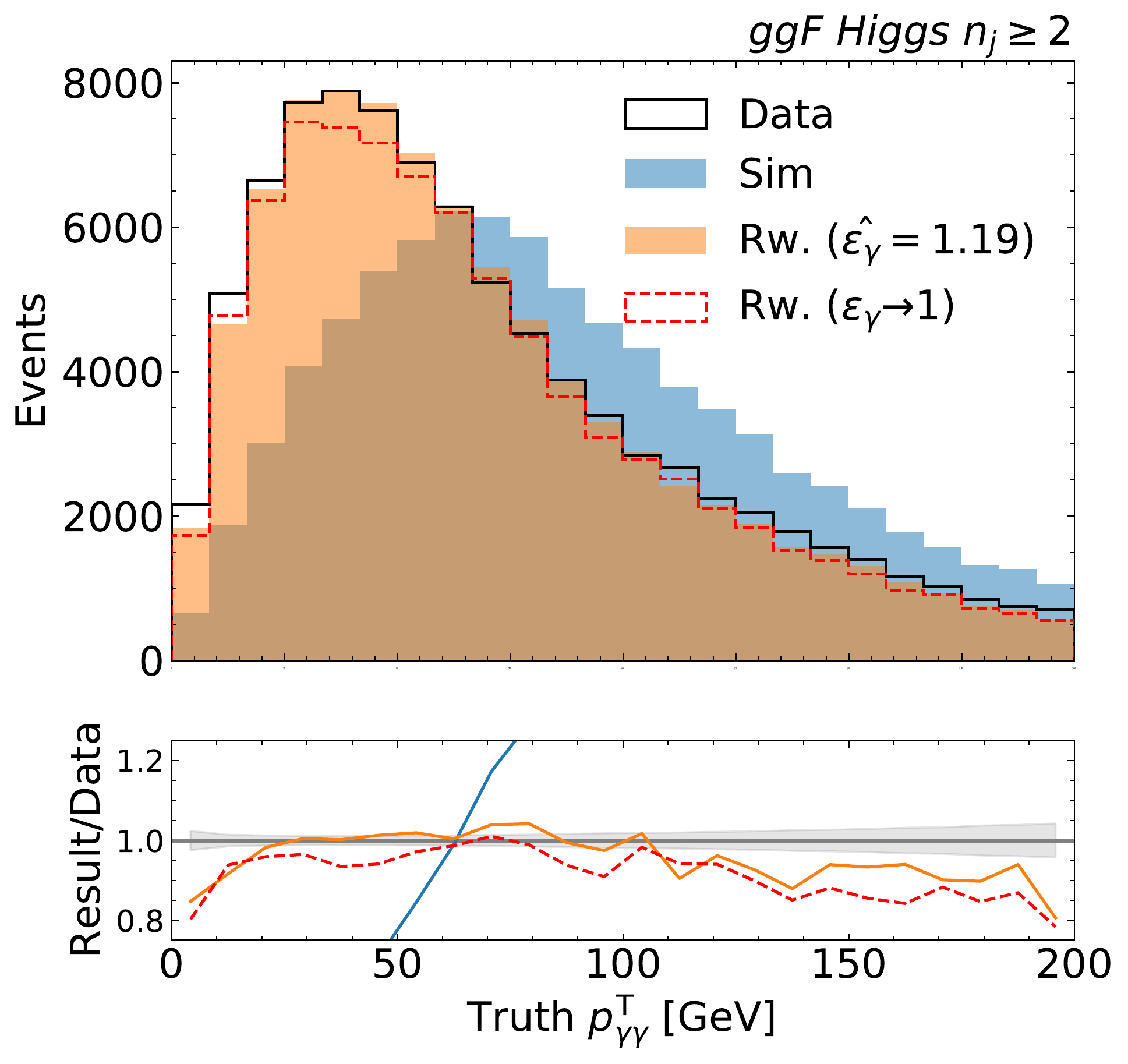}
\caption{Higgs boson cross section: results of the $w_0$ optimization. The nuisance parameter $\epsilon_\gamma$ is optimized simultaneously with $w_0$ with the prior constraint set to 50\% (orange) or fixed to 1 for comparison (red). The fitted $\epsilon_\gamma$ is $1.19 \pm 0.007$. (Top-left) The detector-level spectrum $m_{\gamma\gamma}$ of the simulation template $D_\mathrm{sim}$ reweighted by the trained $w_0 \times w_1$, compared to the $m_{\gamma\gamma}$ spectrum of the observed data $D_\mathrm{obs}$. (Top-right) The detector-level spectrum $p^\mathrm{T}_{\gamma\gamma}$ of the simulation template $D_\mathrm{sim}$ reweighted by the trained $w_0 \times w_1$, compared to the $p^\mathrm{T}_{\gamma\gamma}$ spectrum of the observed data $D_\mathrm{obs}$. (Bottom) The particle-level spectrum $p^\mathrm{T}_{\gamma\gamma}$ of the simulation template $D_\mathrm{sim}$ reweighted by the trained $w_0$, compared to the $p^\mathrm{T}_{\gamma\gamma}$ spectrum of the observed data $D_\mathrm{obs}$. The shaded band in the bottom panel represents the data statistical uncertainty, which is estimated as $1/\sqrt{n}$, where $n$ is the number of observed events in a given bin.}
\label{fig:higgs_w0_theta}
\end{center}
\end{figure*}

The $w_0$ reweighter and $\epsilon$ are optimized simultaneously based on the pre-trained $w_1$ reweighter. The prior of $\epsilon_\gamma$ is 50\%. The fitted $\epsilon_\gamma$ is $1.19 \pm 0.007$. As shown in Fig.~\ref{fig:higgs_w0_theta}, the reweighted spectra match well with observed data in both detector and particle level. This means that the observed data $p^\mathrm{T}_{H}$ spectrum is successfully unfolded with nuisance parameter $\epsilon_{\gamma}$ properly profiled. For comparison, we also perform UPU with $\epsilon_\gamma$ fixed at 1. As shown in Fig.~\ref{fig:higgs_w0_theta}, the unfolded $p^\mathrm{T}_{H}$ spectrum in this case has a larger non-closure with the observed data due to the lack of profiling.


\section{Conclusion and Outlook}
\label{sec:conclusion}

In this paper, we proposed Unbinned Profiled Unfolding (UPU), a new ML-based unfolding method that can process unbinned data and profile. The method uses the binned maximum likelihood as the figure of merit to optimize the unfolding reweighting function $w_0\left(t\right)$, which takes unbinned particle-level spectra as inputs. $w_0\left(t\right)$ and the nuisance parameters $\theta$ are optimized simultaneously, which also requires to learn a conditional likelihood ratio $w_1(t,r|\theta)$ that reweights the detector-level spectra based on the profiled values of nuisance parameters and is taken as an input for the optimization of $w_0\left(t\right)$ and $\theta$.

In the Gaussian example, we demonstrated the optimization of $w_1$ and the optimization of $w_0$ and $\theta$. The setup considers one dimension in the particle level and two dimension in the detector level. The additional detector-level observable which does not depend on $\theta$ breaks the degeneracy between particle-level and detector-level effects and thus allows for optimization of $w_1$ and $\theta$ at the same time.

We also applied UPU to the Higgs boson cross section measurement. We considered one dimension at particle level and two dimensions at detector level. With one detector-level variable sensitive to the target particle-level observable and one sensitive to the effect of nuisance parameters, the data are successfully unfolded and profiled. The impact of profiling is also demonstrated by comparing with the result of nuisance parameter fixed to the nominal value. This can be readily extended to higher dimensions in either particle level or detector level, provided all particle-level and detector-level effects are distinguishable in the considered detector-level spectra. In the case of more than one nuisance parameters, one can either train multiple $w_1$ for each nuisance parameter separately or train a single $w_1$ which takes all nuisance parameters as inputs.  As the effects of multiple nuisance parameters are usually assumed independent, one could take a product of individually trained reweighters.

As with any measurement, quantifying the uncertainty is critical to interpret UPU results.  Just as in the binned case, one can calculate the uncertainty on the nuisance parameters which can be determined by fixing a given parameter to target values and then simultaneously re-optimizing $w_0$ and the rest of the nuisance parameters.  A new feature of UPU is that the likelihood (ratio) itself is only an approximation, using neural networks as surrogate models.  This is a challenge for all machine learning-based unfolding, and uncertainties can be probed by comparing the results with different simulations.  Future extensions of UPU may be able to also use machine learning to quantify these model uncertainties as well as process unbinned data also at detector level.

\section*{Code and data}

\noindent The code for this paper can be found at \href{https://github.com/qwerasd903/UnbinnedProfiledUnfolding}{https://github.com/qwerasd903/UnbinnedProfiledUnfolding}, which uses Jupyter notebooks \cite{Kluyver2016jupyter} and employs NumPy \cite{harris2020array} for data manipulation and Matplotlib \cite{Hunter:2007} for visualization. All of the machine learning was performed on an NVIDIA A100 Graphical Processing Unit (GPU) and reproducing the entire notebook takes about 13 hours. The physics data sets are hosted on Zenodo at Ref.~\cite{chan_jay_2023_7553271}.

\begin{acknowledgments}

We thank A. Ghosh, V. Mikuni, J. Thaler, D. Whiteson and S. L. Wu for helpful discussions and detailed feedback on the manuscript.  BN is supported by the U.S. Department of Energy (DOE), Office of Science under contract DE-AC02-05CH11231. JC is supported by the U.S. Department of Energy (DOE), Office of Science under contracts DE-AC02-05CH11231 and DE-SC0017647.

\end{acknowledgments}

\begin{appendix}

  \section{Gaussian example with one-dimension in both particle and detector level}
\label{sec:ssec:gaussian1D}

In this appendix, we apply UPU to the Gaussian example where each data set represents one-dimensional Gaussian random variables in both the particle and detector level. The particle-level random variable $T$ is described by mean $\mu$ and standard deviation $\sigma$, while the detector-level variable is given by
\begin{equation}
    R = T + Z,
\end{equation}
where $Z$ is a Gaussian random variable with mean $\beta$ and standard deviation $\epsilon$.

$\epsilon$ is considered to be the only nuisance parameter, and $\beta$ is fixed to 0. Similar to the setup in Sec.~\ref{sec:gaussian}, three data sets are prepared for the full training procedure:
\begin{itemize}
    \item $D_\mathrm{sim}^{1.0}$: 200,000 events with $\mu=0$, $\sigma=1$ and $\epsilon=1$
    \item $D_\mathrm{obs}$: 100,000 events with $\mu=0.2$, $\sigma=1$ and $\epsilon=1.2$
    \item $D_\mathrm{sim}^*$: 200,000 events with $\mu=0$, $\sigma=1$ and $\epsilon$ uniformly distributed from 0.2 to 1.8.
    \item $D_\mathrm{sim}^{1.2}$: 100,000 events with $\mu=0$, $\sigma=1$ and $\epsilon=1.2$.
\end{itemize}

\begin{figure}[htbp]
\begin{center}
\includegraphics[width=0.43\textwidth]{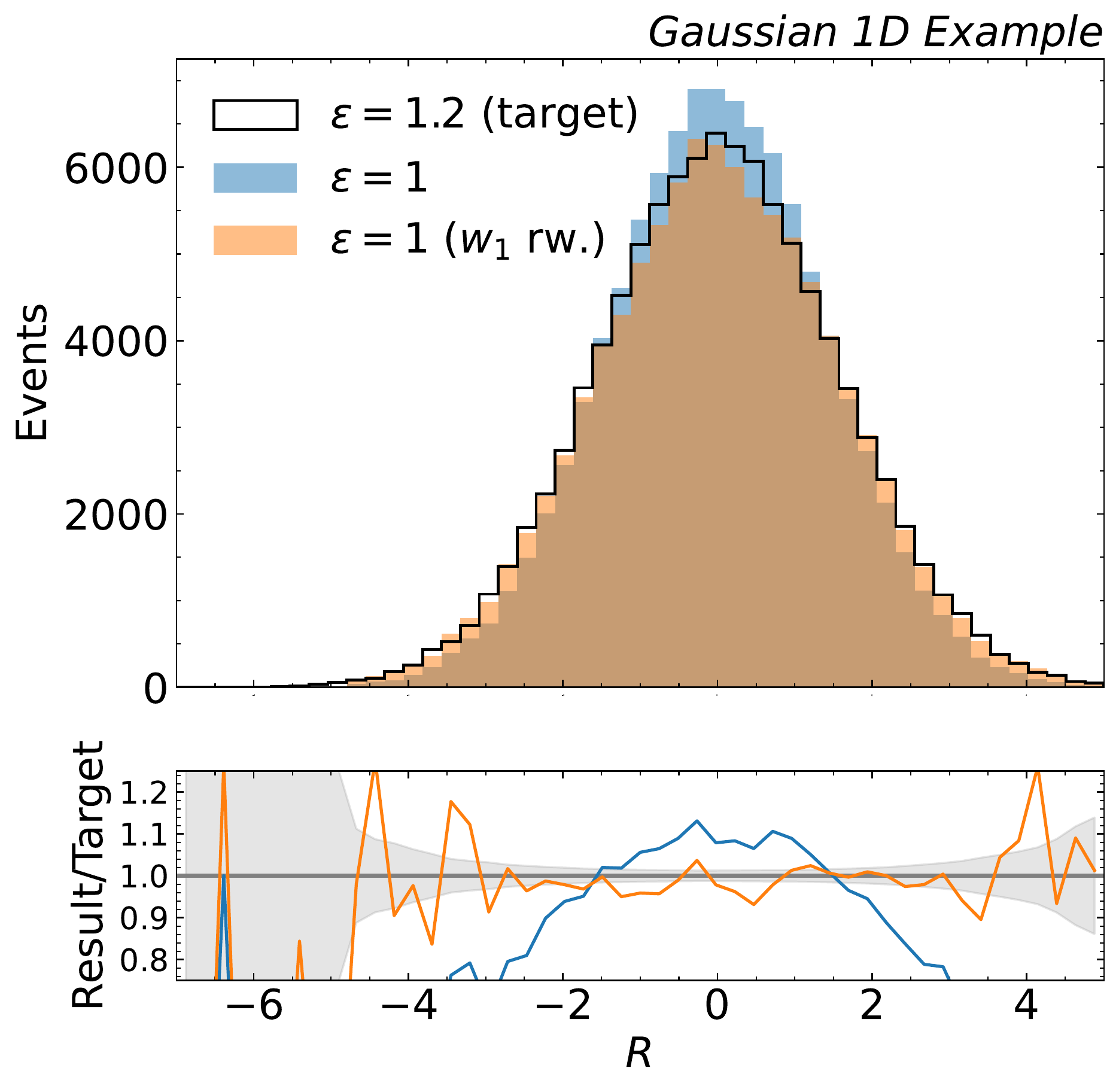}
\caption{Gaussian 1D example: the nominal $R$ distribution ($\epsilon = 1$) reweighted by the trained $w_1$ conditioned at $\epsilon = 1.2$ and compared to $R$ distribution with $\epsilon = 1.2$. The shaded band in the bottom panel represents the data statistical uncertainty, which is estimated as $1/\sqrt{n}$, where $n$ is the number of $D_\mathrm{sim}^{1.2}$ events in a given bin.}
\label{fig:gaussian1D_w1}
\end{center}
\end{figure}

A $w_1$ reweighter is trained to reweight $D_\mathrm{sim}^{1.0}$ to $D_\mathrm{sim}^*$. The trained $w_1$ is then tested with the nominal $R$ distribution ($D_\mathrm{sim}^{1.0}$) reweighted to $\epsilon = 1.2$ ($w_1\left(R|T,\epsilon=1.2\right)$) and compared to the $R$ spectrum with $\epsilon = 1.2$ ($D_\mathrm{sim}^{1.2}$). As shown in Fig. \ref{fig:gaussian1D_w1}, the trained $w_1$ reweighter has learned to reweight the nominal $R$ spectrum to match the $R$ spectrum with $\epsilon$ at 1.2.

\begin{figure}[htbp]
\begin{center}
\includegraphics[width=0.43\textwidth]{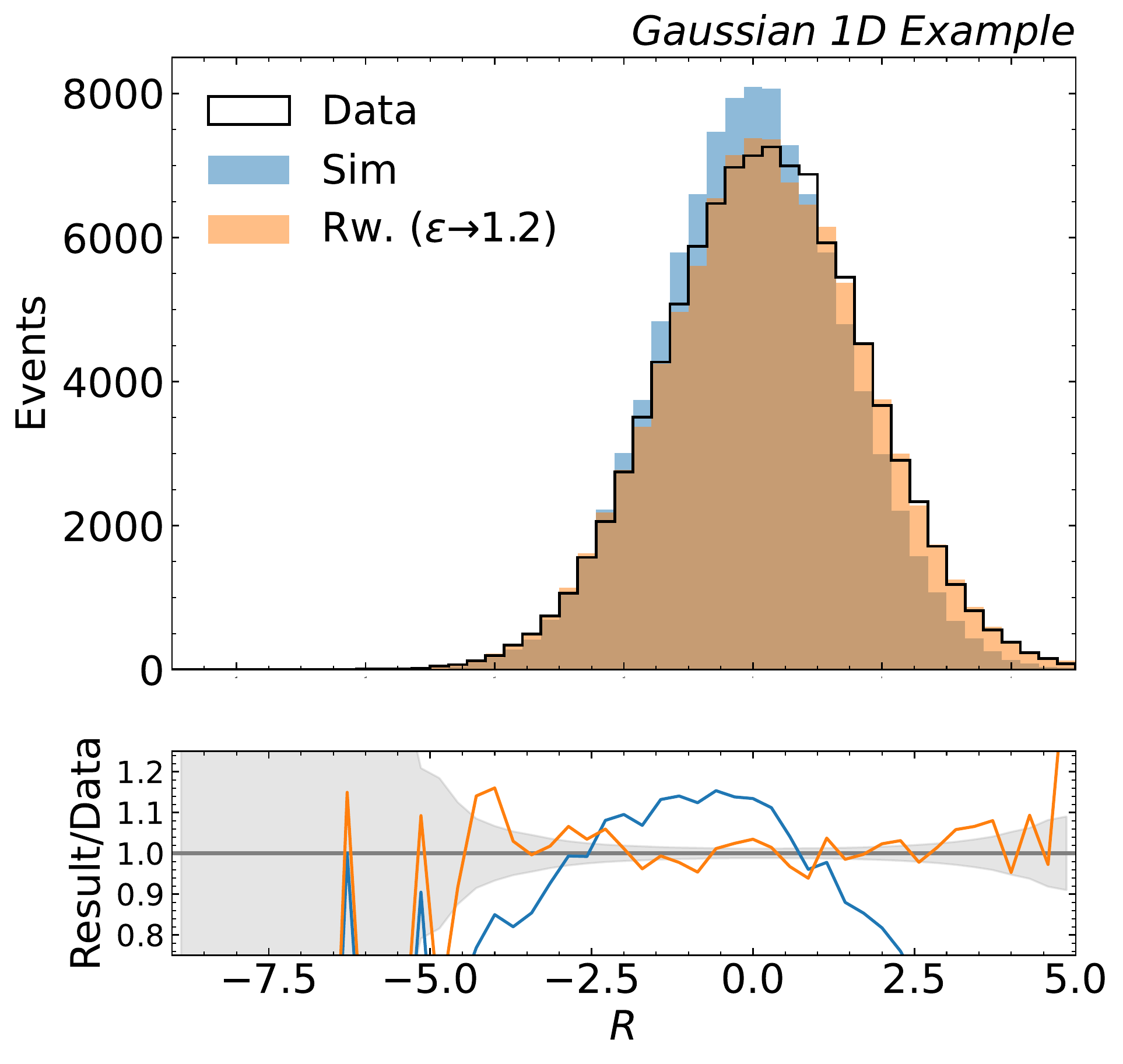}
\includegraphics[width=0.43\textwidth]{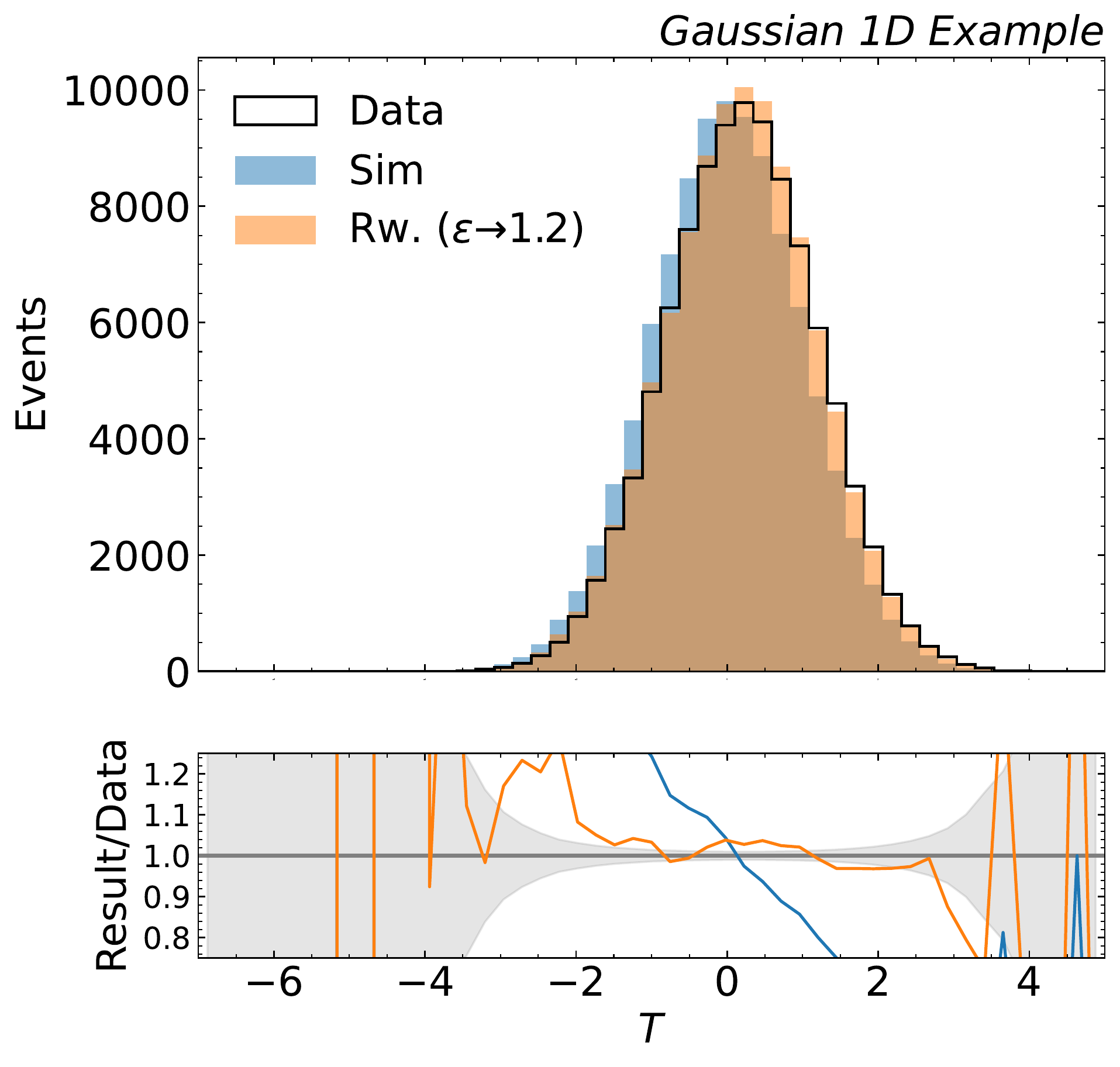}
\caption{Gaussian 1D example: results of the $w_0$ optimization. The nuisance parameter $\epsilon$ is fixed to 1.2, and the the penalty term in Eq.~\ref{eq:lllall} is set to 0. (Top) The detector-level spectrum $R$ of the simulation template $D_\mathrm{sim}$ reweighted by the trained $w_0 \times w_1$, compared to the $R$ spectrum of the observed data $D_\mathrm{obs}$. (Bottom) The particle-level spectrum $T$ of the simulation template $D_\mathrm{sim}$ reweighted by the trained $w_0$, compared to the $T$ spectrum of the observed data $D_\mathrm{obs}$. The shaded band in the bottom panel represents the data statistical uncertainty, which is estimated as $1/\sqrt{n}$, where $n$ is the number of observed events in a given bin.}
\label{fig:gaussian1D_w0_noepsilon}
\end{center}
\end{figure}

\begin{figure*}[htbp]
\begin{center}
\includegraphics[width=0.43\textwidth]{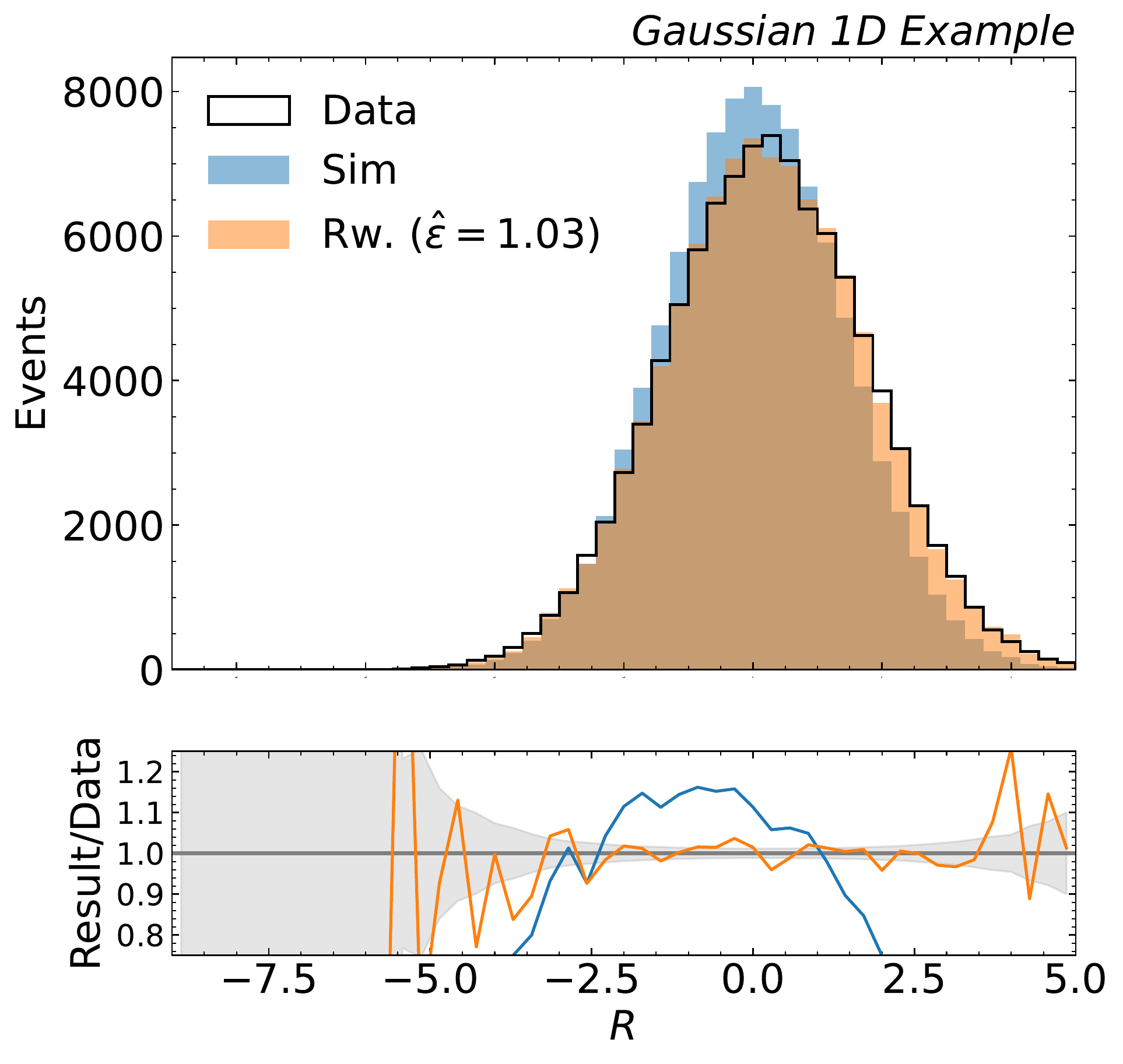}
\includegraphics[width=0.43\textwidth]{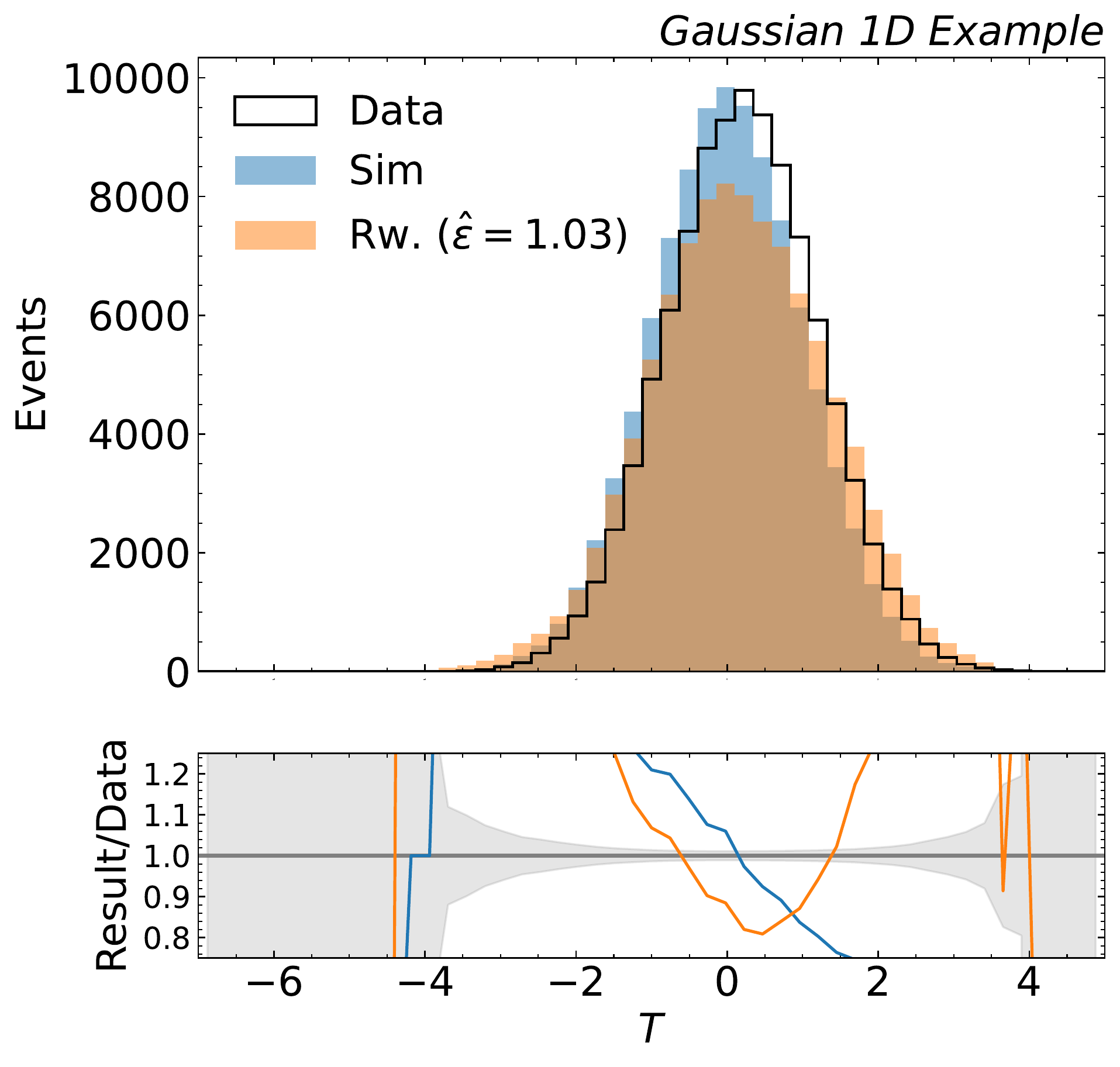}
\caption{Gaussian 1D example: results of the $w_0$ optimization. The nuisance parameter $\epsilon$ is optimized simultaneously with $w_0$ and the best-fit value is $\hat{\epsilon} = 1.03 \pm 0.016$. (Left) The detector-level spectrum $R$ of the simulation template $D_\mathrm{sim}$ reweighted by the trained $w_0 \times w_1$, compared to the $R$ spectrum of the observed data $D_\mathrm{obs}$. (Right) The particle-level spectrum $T$ of the simulation template $D_\mathrm{sim}$ reweighted by the trained $w_0$, compared to the $T$ spectrum of the observed data $D_\mathrm{obs}$. The shaded band in the bottom panel represents the data statistical uncertainty, which is estimated as $1/\sqrt{n}$, where $n$ is the number of observed events in a given bin.}
\label{fig:gaussian1D_w0_epsilon}
\end{center}
\end{figure*}

With this trained $w_1$ reweighter, a $w_0$ reweighter is trained using $D_\mathrm{sim}^{1.0}$ as the simulation template with $D_\mathrm{obs}$ as the observed data used in Eq.~\ref{eq:lllall}. In the first scenario, the nuisance parameter $\epsilon$ for the $w_1$ reweighter is fixed to 1.2, and the penalty term in Eq.~\ref{eq:lllall} $\log(\theta)$ is set to 0 (no constraint). As shown in Fig.~\ref{fig:gaussian1D_w0_noepsilon}, the $w_0$ reweighter is able to learn to reweight the particle-level spectrum $T$ by matching the detector-level spectrum $R$ to the observed spectrum. In the second scenario, the nuisance parameter $\epsilon$ is trained together with the $w_0$ reweighter. The prior in the penalty term in Eq.~\ref{eq:lllall} is set to be a Gaussian probability density with a 80\% uncertainty. As shown in Fig. \ref{fig:gaussian1D_w0_epsilon}, the trained $w_0$ and optimized $\epsilon$ are tested. The fitted $\epsilon$ is $1.03 \pm 0.016$ \footnote{The fitted value is averaged over five different $w_0$ reweighters which are trained in the same way, but with different random initializations. The standard deviation of the fitted values is taken as the error.} (true value is 1.2). The reweighted distribution matches well with observed data in the detector-level spectrum but the particle-level spectrum has a large non-closure. This is because of the degeneracy between the $w_0$ and $w_1$ reweighters in the effect on the detector-level spectrum.  In other words, detector effects can mimic changes in the particle-level cross section, so the data cannot distinguish between these two scenarios. This is a common issue which also exists in the standard binned maximum likelihood unfolding. For comparison, we also perform the standard binned maximum likelihood unfolding. As shown in App.~\ref{app:bmlu}, the unfolded $T$ spectrum in this case also fails to represent the true $T$ spectrum. An 80\% uncertainty is highly exaggerated from typical scenarios, but it clearly illustrates the challenge of profiling and unfolding at the same time.

  \section{Binned maximum likelihood unfolding with Gaussian examples}
  \label{app:bmlu}
  
\begin{figure*}[htbp]
\begin{center}
\includegraphics[width=0.43\textwidth]{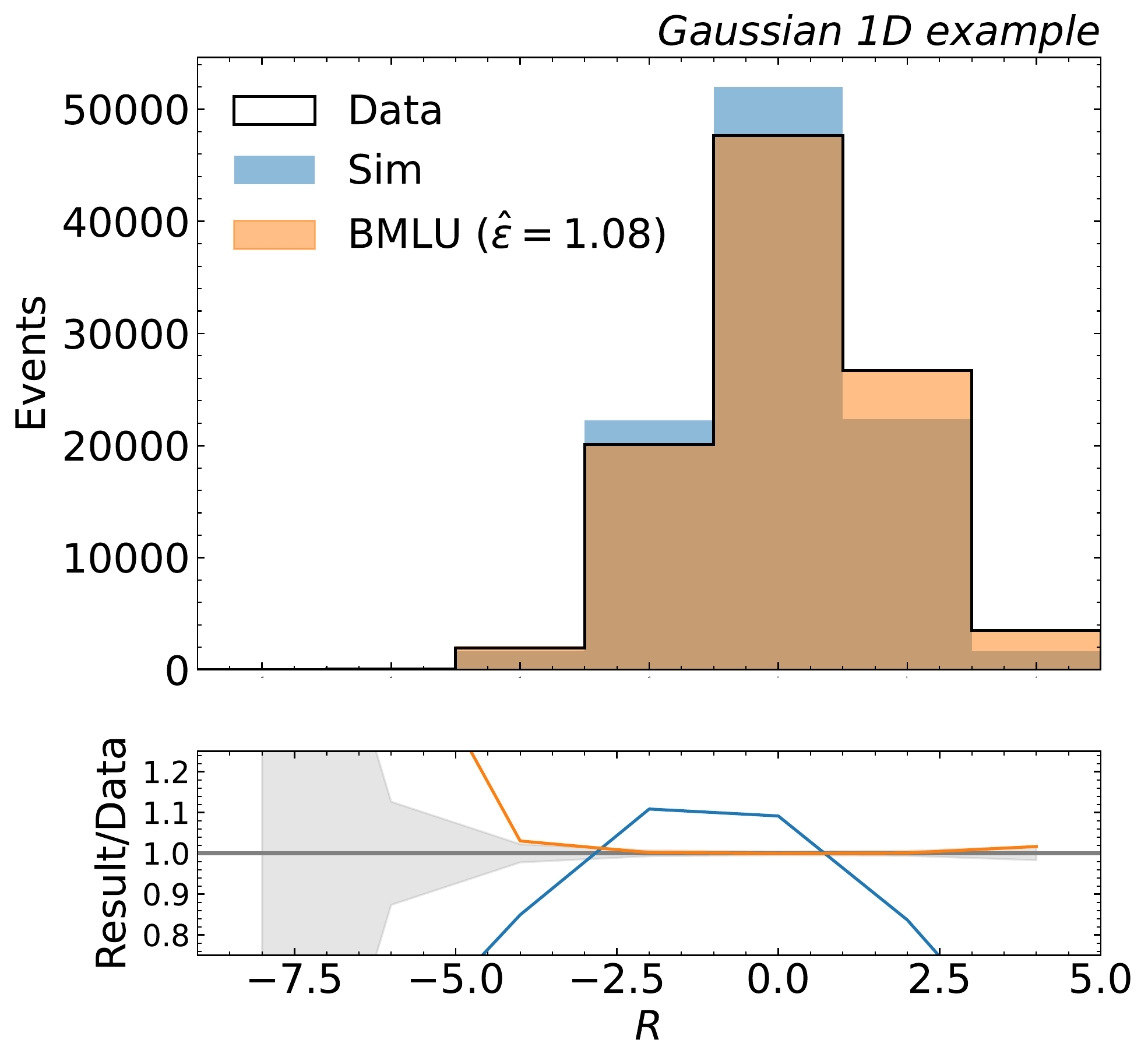}
\includegraphics[width=0.43\textwidth]{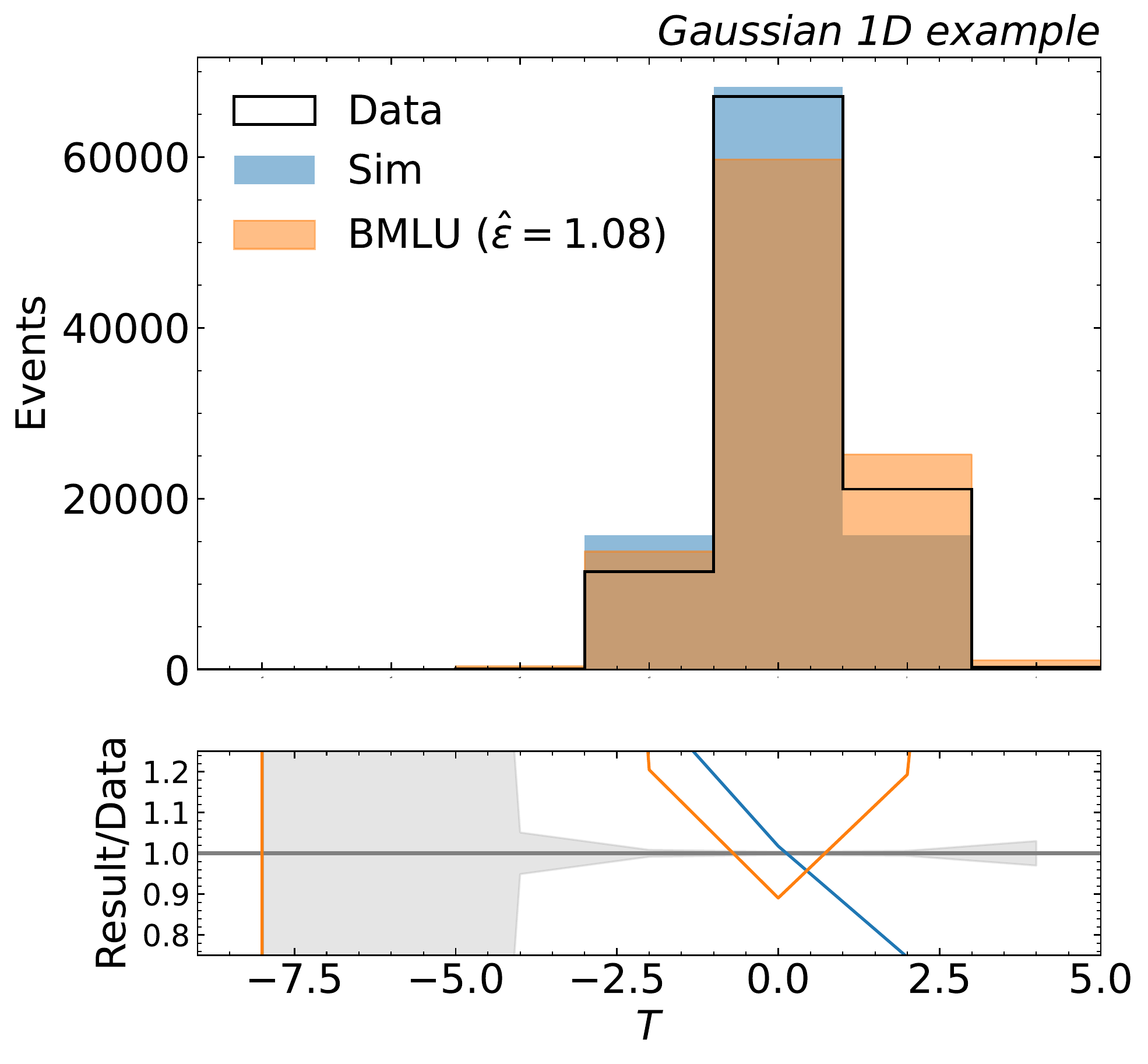}
\caption{Gaussian 1D example: results of the binned maximum likelihood unfolding. The prior constraint for $\epsilon$ is set to 80\% and the fitted $\epsilon$ is $1.08 \pm 0.02$. (Left) The fitted detector-level spectrum $R$ of the simulation template $D_\mathrm{sim}$, compared to the $R$ spectrum of the observed data $D_\mathrm{obs}$. (Right) The unfolded particle-level spectrum $T$ of the simulation template $D_\mathrm{sim}$, compared to the $T$ spectrum of the observed data $D_\mathrm{obs}$. The shaded band in the bottom panel represents the data statistical uncertainty, which is estimated as $1/\sqrt{n}$, where $n$ is the number of observed events in a given bin.}
\label{fig:gaussian1D_bmlu}
\end{center}
\end{figure*}

\begin{figure*}[htbp]
\begin{center}
\includegraphics[width=0.43\textwidth]{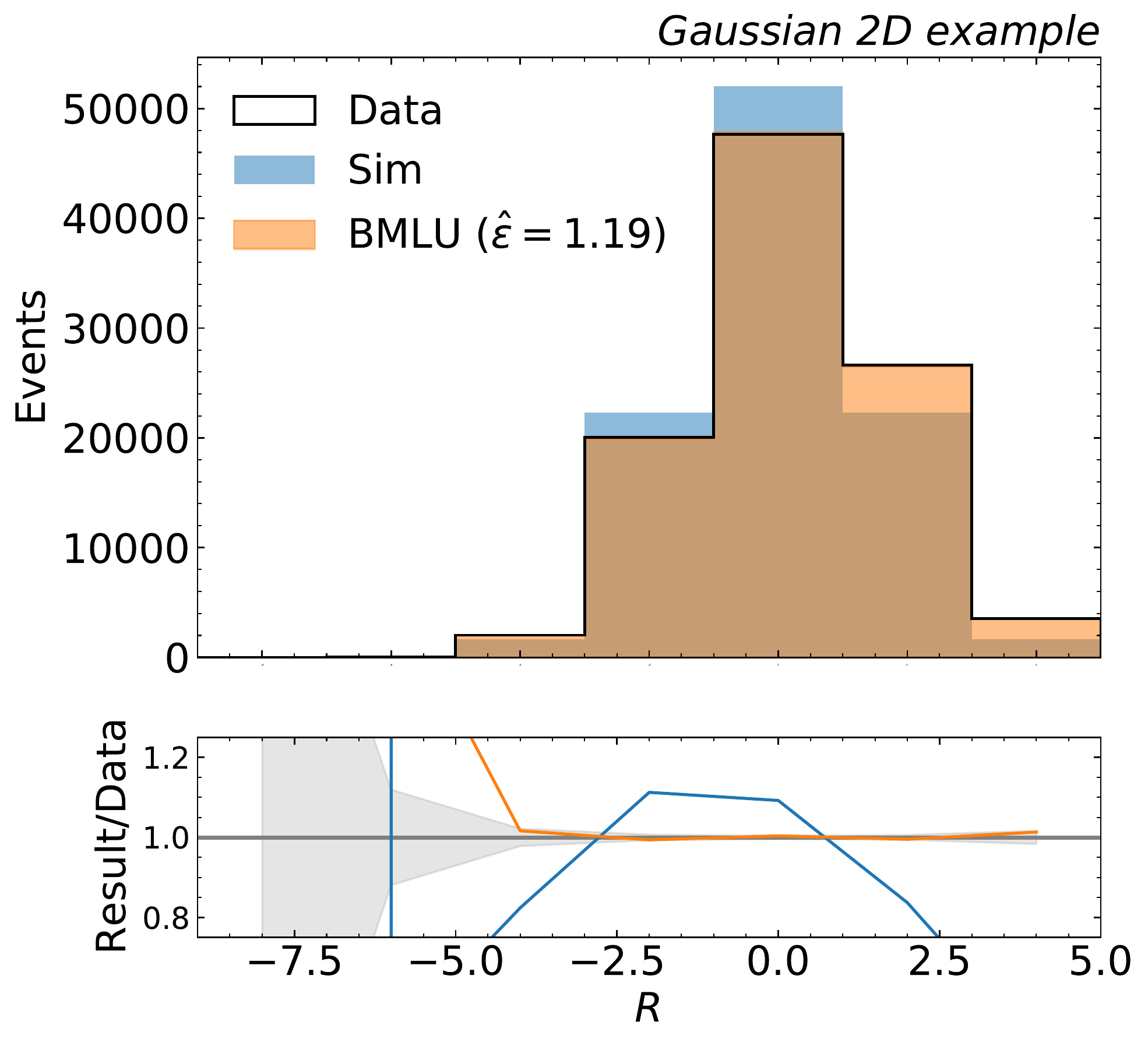}
\includegraphics[width=0.43\textwidth]{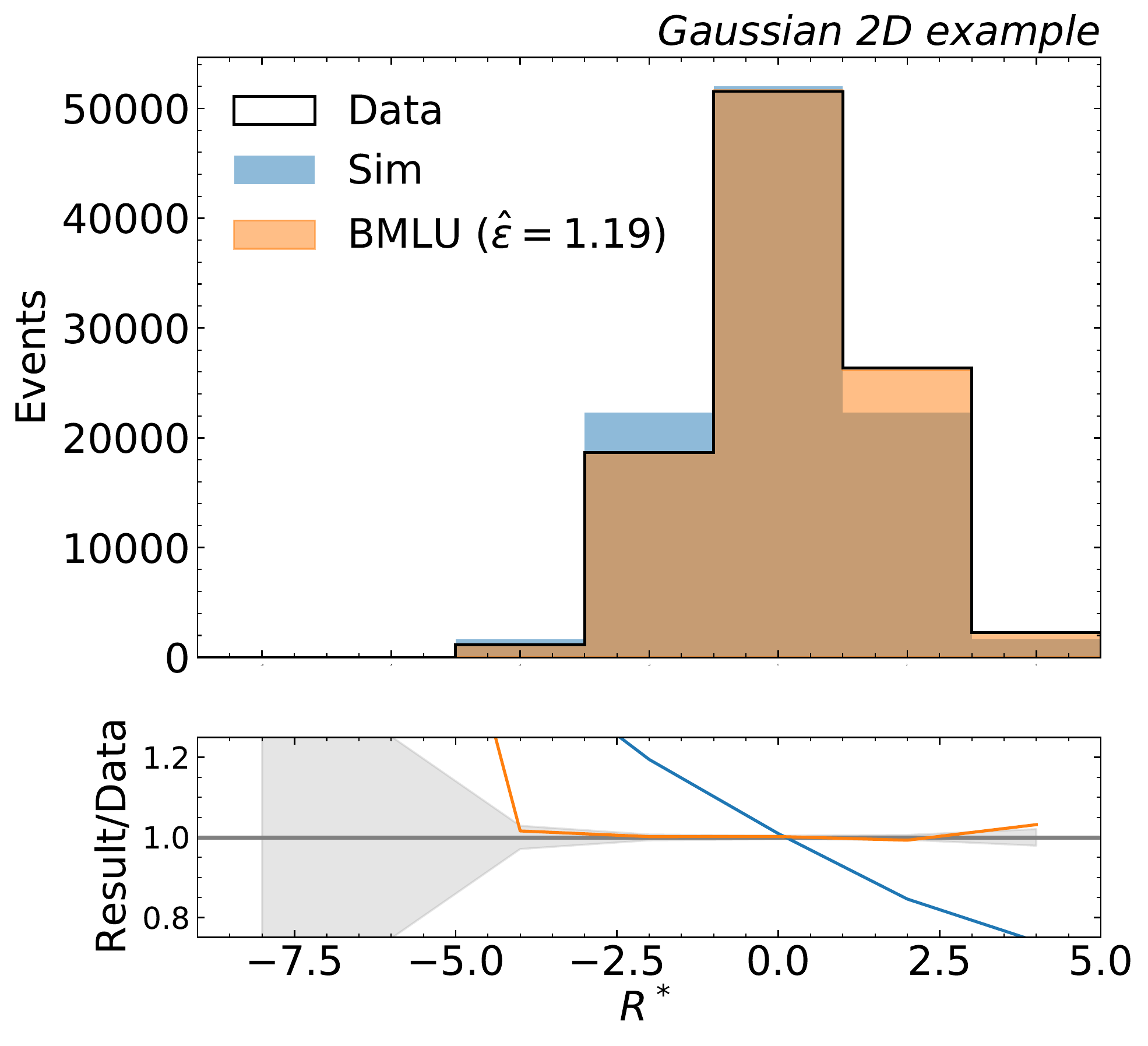}\\
\includegraphics[width=0.43\textwidth]{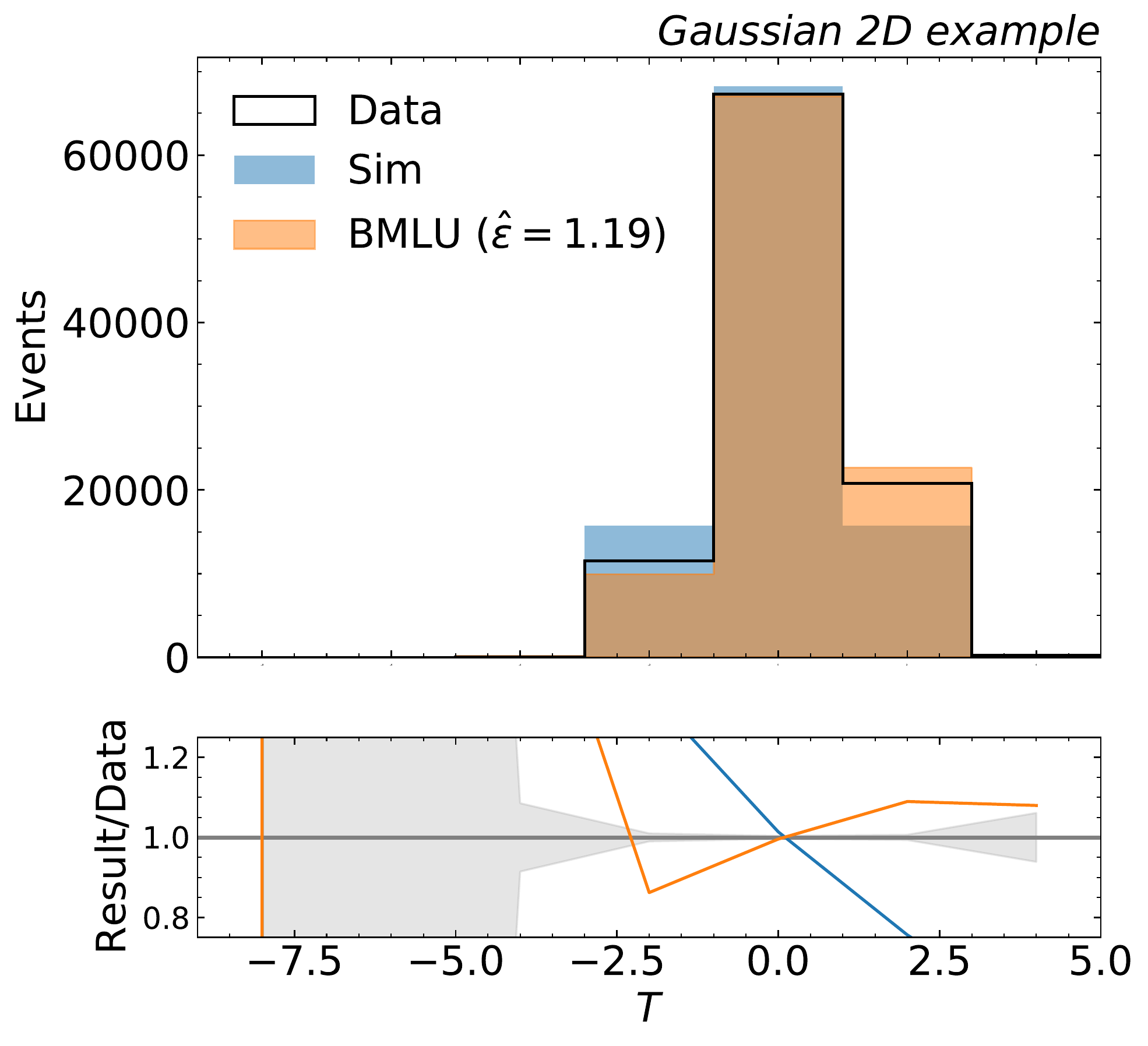}
\caption{Gaussian 2D example: results of the binned maximum likelihood unfolding. The prior constraint for $\epsilon$ is set to 80\% and the fitted $\epsilon$ is $1.19 \pm 0.003$. (Top-left) The fitted detector-level spectrum $R$ of the simulation template $D_\mathrm{sim}$, compared to the $R$ spectrum of the observed data $D_\mathrm{obs}$. (Top-right) The fitted detector-level spectrum $R^*$ of the simulation template $D_\mathrm{sim}$, compared to the $R^\prime$ spectrum of the observed data $D_\mathrm{obs}$. (Bottom) The unfolded particle-level spectrum $T$ of the simulation template $D_\mathrm{sim}$, compared to the $T$ spectrum of the observed data $D_\mathrm{obs}$. The shaded band in the bottom panel represents the data statistical uncertainty, which is estimated as $1/\sqrt{n}$, where $n$ is the number of observed events in a given bin.}
\label{fig:gaussian2D_bmlu}
\end{center}
\end{figure*}

  In this appendix, we present results of the standard binned maximum likelihood unfolding (BMLU) with Gaussian examples. The scenarios are:
  
  \begin{itemize}
      \item One-dimension in both particle and detector
level: this is the same example as described in App.~\ref{sec:ssec:gaussian1D}. The prior constraint for $\epsilon$ is set to 80\%. The result is shown in Fig.~\ref{fig:gaussian1D_bmlu} with $\epsilon$ fitted to $1.08 \pm 0.02$, which also indicates a degeneracy problem between particle and detector levels.
       \item One-dimension in particle level and
two-dimension in detector level: this is the same example as described in Sec.~\ref{sec:gaussian}. The prior constraint for $\epsilon$ is set to 80\%. The result is shown in Fig.~\ref{fig:gaussian2D_bmlu} with $\epsilon$ fitted to $1.19 \pm 0.003$. The degeneracy problem is resolved after considering an additional spectrum in the detector level.

All the maximum likelihood fittings are performed using pyhf \cite{pyhf, pyhf_joss}.
  \end{itemize}
  
\end{appendix}

 \bibliography{mybib}

\end{document}